\documentclass[12pt]{article}

\usepackage{bbm,latexsym}
\usepackage{amssymb,amsmath}
\usepackage{graphicx}
\usepackage[font=small,labelfont=bf]{caption}
\usepackage[font=small,labelfont=small]{subcaption}
\usepackage{slashed}
\usepackage{feynmp-auto}
\newcommand{\mc}{\mathcal{M}}
\newcommand{\diag}{\mbox{diag}}
\newcommand{\bone}{\mathbbm{1}}
\newcommand{\bb}{{\scriptscriptstyle (B)}}
\newcommand{\dd}{\mathrm{d}}

\newcommand{\clam}[1]{\left( #1 \right)}
\allowdisplaybreaks
\begin{document}

\title{
\normalsize \hfill UWThPh-2017-11 \\[10mm]
\LARGE Renormalization and radiative corrections to masses \\
in a general Yukawa model} 

\author{
M.~Fox\thanks{E-mail: a0905865@unet.univie.ac.at}\;,
\addtocounter{footnote}{1}
W.~Grimus\thanks{E-mail: walter.grimus@univie.ac.at}\;
and M.~L\"oschner\thanks{E-mail: maximilian.loeschner@univie.ac.at}
\\[5mm]
\small University of Vienna, Faculty of Physics \\
\small Boltzmanngasse 5, A--1090 Vienna, Austria
}

\date{October 2, 2017}

\maketitle

\begin{abstract}
We consider a model with arbitrary numbers of Majorana fermion fields and real
scalar fields $\varphi_a$, general Yukawa couplings and 
a $\mathbbm{Z}_4$ symmetry that forbids linear and trilinear terms 
in the scalar potential. Moreover, fermions become massive only after 
spontaneous symmetry breaking of the $\mathbbm{Z}_4$ symmetry 
by vacuum expectation
values (VEVs) of the $\varphi_a$. Introducing the shifted fields $h_a$ whose
VEVs vanish, $\overline{\mbox{MS}}$ renormalization
of the parameters of the unbroken theory suffices to make the theory
finite. However, in this way, beyond tree level it is necessary to perform
finite shifts of the tree-level VEVs, induced by the finite parts of the
tadpole diagrams, in order to ensure vanishing one-point functions of the
$h_a$. 
Moreover, adapting the renormalization scheme to a situation with many scalars 
and VEVs, we consider the \emph{physical} fermion and scalar masses as derived 
quantities, \textit{i.e.}\ as functions of the coupling constants and VEVs.   
Consequently, the masses have to be computed order by order in a 
perturbative expansion.
In this scheme we compute the selfenergies of 
fermions and bosons and show how to obtain the respective one-loop 
contributions to the tree-level masses.
Furthermore, we discuss the modification of our results in the case
of Dirac fermions and investigate, by way of an example, the effects of a
flavour symmetry group.
\end{abstract}

\newpage

\section{Introduction}
\label{introduction}
Thanks to the results of the neutrino oscillation experiments---see for 
instance~\cite{rpp,nuosc-exp}---it is now 
firmly established that at least two light neutrinos have a nonzero mass
and that there is a non-trivial lepton mixing matrix or PMNS matrix in analogy
to the quark mixing matrix or CKM matrix. 
The surprisingly large mixing angles in 
the PMNS matrix have given a boost to model building with spontaneously
broken flavour symmetries---for recent reviews see~\cite{reviews}. 
Many interesting results
have been discovered, however, no favoured scenario has emerged yet. 
Moreover, predictions of neutrino mass and mixing models refer frequently to
tree-level computations. It would thus be desirable to check the stability of
such predictions under radiative corrections. In the case of renormalizable
models one has a clear-cut and consistent method to remove ultraviolet 
(UV) divergences and to compute such corrections.

However, there is the complication that the envisaged models always have a
host of scalars and often complicated spontaneous symmetry breaking (SSB) of
the flavour group. This makes it impossible to replace all
Yukawa couplings by ratios of masses over vacuum expectation values (VEVs), 
as done for instance in the renormalization of the Standard Model. 
Of course, one could replace part
of the Yukawa coupling constants by masses, but this would make the
renormalization procedure highly asymmetric. In this paper we suggest to
make such models finite by $\overline{\mbox{MS}}$ 
renormalization of the parameters of the unbroken
model and to perform finite VEV shifts at the loop level in order to guarantee
vanishing  scalar one-point functions of the shifted 
scalar fields~\cite{weinberg}. 
Additionally, we introduce finite field strength renormalization for 
obtaining on-shell selfenergies. 
In this way, all fermion and scalar masses 
are derived quantities and functions of the parameters of the model.

In the usual approach to renormalization of theories with SSB and 
mixing~\cite{Aoki:1982} one has counterterms for masses, 
quark and lepton mixing matrices---see for 
instance~\cite{Denner:1990,Kniehl:1996}---and tadpoles---see for 
instance~\cite{Fleischer:1980ub,Denner:2016etu}.\footnote{There are other 
treatments of tadpoles adapted to the theory where they occur, 
see for instance reference~\cite{Pierce:1993} for the MSSM
and~\cite{Sperling:2013eva,Krause:2016oke} where the issue of gauge 
invariance is discussed.}
We stress that in our approach there are no such counterterms
because we use an alternative approach tailored to the situation with a
proliferation of scalars and VEVs.

In order to present the renormalization scheme in a clear and compact way,
we consider a toy model 
which has 
\begin{itemize}
\item
an arbitrary number of Majorana or Dirac fermions,
\item
an arbitrary number of neutral scalars,
\item
a $\mathbbm{Z}_4$ ($\mathbbm{Z}_2$) symmetry which forbids 
Majorana (Dirac) fermion masses before 
SSB\footnote{This is motivated by the Standard Model 
where---before SSB---fermion masses 
as well as linear and trilinear terms in the scalar potential 
are absent due to the gauge symmetry.} 
and 
\item
general Yukawa interactions.
\end{itemize}
We put particular emphasis on the treatment of tadpoles. Since radiative 
corrections in this model are already finite due to
$\overline{\mbox{MS}}$ renormalization with the counterterms of the 
unbroken theory, also the sum of all tadpole contributions, \textit{i.e.}\ 
the loop contributions and those induced by the counterterms of the unbroken 
theory, is finite. 
However, tadpoles introduce finite VEV shifts which have to be taken 
into account for instance in the selfenergies. Eventually, the finite 
VEV shifts also contribute to the radiative corrections of the tree-level 
masses.\footnote{After SSB, these shifts have to be taken into account
everywhere in the Lagrangian where VEVs appear in order to obtain a consistent
set of counterterms.}
We also focus on Majorana fermions, having in mind that neutrinos automatically 
obtain Majorana nature through the seesaw mechanism~\cite{seesaw}.

An attempt at a renormalization 
scheme---with one fermion and one scalar field---along the lines discussed here 
has already been made in~\cite{Grimus:2014zwa}; however, 
the treatment of the VEV in this paper cannot be generalized to the case of
more than one scalar field. 

The paper is organized as follows. In section~\ref{setup} we introduce 
the Lagrangian, define the counterterms and discuss SSB. 
Section~\ref{renormalization} is devoted to the explanation of our 
renormalization scheme, while in section~\ref{one-loop} we explicitly compute 
the selfenergies of fermions and scalars at one-loop order. We present an 
example of a flavour symmetry in section~\ref{example} and study how the 
symmetry teams up with the general renormalization scheme.
In section~\ref{dirac} we describe the changes when one has Dirac fermions 
instead of Majorana fermions. Finally, in section~\ref{conclusions} 
we present the conclusions. 
Some details which are helpful for reading the paper 
can be found in the three appendices.

\section{Toy model setup}
\label{setup}
In this section, we give the specifics of the investigated model and 
discuss the generation of masses via SSB. We focus on Majorana fermions.
Throughout this paper we always use the sum convention, if not otherwise stated.

\subsection{Bare and renormalized Lagrangian}
\label{lagrangian}
The bare Lagrangian is given by
\begin{subequations}\label{eq:general-model}
\begin{eqnarray}
\mathcal{L}_B &=& 
i \bar \chi^\bb_{iL} \gamma^\mu \partial_\mu \chi^\bb_{iL} +
\frac{1}{2} \left( \partial_\mu \varphi^\bb_a \right) 
\left( \partial^\mu \varphi^\bb_a \right)
\\ &&
+ \left( \frac{1}{2}\,\left(Y^\bb_a\right)_{ij}\, 
{\chi^\bb_{iL}}^T C^{-1} \chi^\bb_{jL} \varphi^\bb_a + \mbox{H.c.} \right) 
\\ &&
- \frac{1}{2} ( \mu_{\scriptscriptstyle B}^2 )_{ab} \varphi^\bb_a \varphi^\bb_b -
\frac{1}{4} \lambda^\bb_{abcd}\, 
\varphi^\bb_a \varphi^\bb_b \varphi^\bb_c \varphi^\bb_d. \label{eq:V-bare}
\end{eqnarray}
\end{subequations}
The charge-conjugation matrix $C$ acts only on the Dirac indices.
We assume $n_\chi$ chiral 
Majorana fermion fields $\chi^\bb_{iL}$ and 
$n_\varphi$ real scalar fields $\varphi^\bb_a$. 
This Lagrangian exhibits the $\mathbbm{Z}_4$ symmetry
\begin{equation}\label{S}
\mathcal{S}: \quad 
\chi^\bb_{L} \to i\chi^\bb_{L}, \quad \varphi^\bb \to -\varphi^\bb,
\end{equation}
with
\begin{equation}
\chi^\bb_{L} = 
\left(
\begin{array}{c} \chi^\bb_{1L} \\ \vdots \\ \chi^\bb_{n_\chi L}
\end{array} \right), \quad
\varphi^\bb = 
\left(
\begin{array}{c} \varphi^\bb_1 \\ \vdots \\ \varphi^\bb_{n_\varphi} 
\end{array} \right).
\end{equation}
Note that
\begin{equation}\label{symm}
\left( Y^\bb_a \right)^T = Y^\bb_a \;\;\forall a = 1,\ldots,n_\chi, \quad
( \mu_{\scriptscriptstyle B}^2 )_{ab} = ( \mu_{\scriptscriptstyle B}^2 )_{ba}
\end{equation}
and $\lambda^\bb_{abcd}$ is symmetric in all indices.\footnote{One can show 
that the number of independent elements of $\lambda^\bb_{abcd}$ is 
$\left( { n_\varphi + 3 \atop 4 } \right)$.} 

We define the renormalized fields by 
\begin{equation}
\chi^\bb_{L} = Z_\chi^{(1/2)}\, \chi_{L}, \quad
\varphi^\bb = Z^{(1/2)}_\varphi\, \varphi,
\end{equation}
where $\chi_{L}$ and $\varphi$ are the vectors of the renormalized fermion and
scalar fields, respectively. The quantity $Z_\chi^{(1/2)}$ is 
a general complex $n_\chi \times n_\chi$ matrix, while 
$Z^{(1/2)}_\varphi$ is a real but otherwise general 
$n_\varphi \times n_\varphi$ matrix.
Since we use dimensional regularization with dimension
\begin{equation}
d = 4 - \varepsilon,
\end{equation}
we also introduce an arbitrary mass parameter $\mc$ which renders 
the renormalized Yukawa and quartic coupling constants dimensionless. 
We split the bare Lagrangian into 
\begin{equation}
\mathcal{L}_B = \mathcal{L} + \delta \mathcal{L},
\end{equation}
where the renormalized Lagrangian is given by 
\begin{subequations}\label{eq:L-Majo}
\begin{eqnarray}
\mathcal{L} &=& 
i \bar \chi_{iL} \gamma^\mu \partial_\mu \chi_{iL} +
\frac{1}{2} \left( \partial_\mu \varphi_a \right) 
\left( \partial^\mu \varphi_a \right)
\\ &&
+ \left( \frac{1}{2}\,\mc^{\varepsilon/2} \left(Y_a\right)_{ij}\, 
\chi_{iL}^T C^{-1} \chi_{jL} \varphi_a + \mbox{H.c.} \right) 
\\ &&
- \frac{1}{2} \mu^2_{ab} \varphi_a \varphi_b -
\frac{1}{4} \mc^{\varepsilon} \lambda_{abcd}\, 
\varphi_a \varphi_b \varphi_c \varphi_d
\end{eqnarray}
\end{subequations}
and 
\begin{subequations}\label{eq:deltaL-Majo}
\begin{eqnarray}
\delta\mathcal{L} &=&
i \delta^{(\chi)}_{ij} \bar \chi_{iL} \gamma^\mu \partial_\mu \chi_{jL} +
\frac{1}{2} \delta^{(\varphi)}_{ab} \left( \partial_\mu \varphi_a \right) 
\left( \partial^\mu \varphi_b \right)
\\ &&
+ \left( \frac{1}{2}\,\mc^{\varepsilon/2} \left(\delta Y_a\right)_{ij}\, 
\chi_{iL}^T C^{-1} \chi_{jL} \varphi_a + \mbox{H.c.} \right) 
\\ \label{V-tree} &&
- \frac{1}{2} \delta\mu^2_{ab} \varphi_a \varphi_b -
\frac{1}{4} \mc^{\varepsilon} \delta\lambda_{abcd}\, 
\varphi_a \varphi_b \varphi_c \varphi_d
\end{eqnarray}
\end{subequations}
contains the counterterms. 
In $\delta \mathcal{L}$, the counterterms corresponding to the 
parameters in $\mathcal{L}$ are given by
\begin{subequations}
\begin{eqnarray}
\label{dY}
\mc^{\varepsilon/2} \delta Y_a &=&
\left( Z_\chi^{(1/2)} \right)^T Y^\bb_b\, Z_\chi^{(1/2)} 
\left( Z^{(1/2)}_\varphi \right)_{ba} - \mc^{\varepsilon/2} Y_a,
\\
\label{dlambda}
\mc^{\varepsilon} \delta\lambda_{abcd} &=& 
\lambda^\bb_{a'b'c'd'} 
\left( Z^{(1/2)}_\varphi \right)_{a'a} 
\left( Z^{(1/2)}_\varphi \right)_{b'b} 
\left( Z^{(1/2)}_\varphi \right)_{c'c} 
\left( Z^{(1/2)}_\varphi \right)_{d'd} 
- \mc^{\varepsilon} \lambda_{abcd},
\\
\label{dmu2}
\delta \mu^2 &=& \left( {Z^{(1/2)}_\varphi} \right)^T \mu^2_{\scriptscriptstyle B} 
\, Z^{(1/2)}_\varphi - \mu^2.
\end{eqnarray}
\end{subequations}
Note that, whenever possible, we use matrix notation, 
as done in equations~(\ref{dY}) and~(\ref{dmu2}).
Moreover we have defined
\begin{equation}\label{delta}
\delta^{(\chi)} = 
\left( Z_\chi^{(1/2)} \right)^\dagger Z_\chi^{(1/2)} - \bone, \quad
\delta^{(\varphi)} = 
\left( Z_\varphi^{(1/2)} \right)^T Z_\varphi^{(1/2)} - \bone.
\end{equation}
The renormalized parameters have the same symmetry properties 
as the unrenormalized ones, \textit{i.e.} 
\begin{equation}
Y_a^T = Y_a \quad \forall a = 1,\ldots,n_\chi, \quad
\mu^2_{ab} = \mu^2_{ba}
\end{equation}
and $\lambda_{abcd}$ is symmetric in all indices. The same applies to the 
corresponding counterterms.

\subsection{Spontaneous symmetry breaking}
\label{SSB}
We introduce the shift 
\begin{equation}\label{shift}
\varphi_a = \mc^{-\varepsilon/2} \bar v_a + h_a
\quad \mbox{with} \quad 
\bar v_a = v_a + \Delta v_a.
\end{equation}
For convenience we have split the shift into $v_a$ and $\Delta v_a$;
below we will identify the $v_a$ with the 
tree-level VEVs of the scalar fields $\varphi_a$,
while the $\Delta v_a$ indicate further 
finite shifts effected by loop corrections.
Throughout our calculations, the symbol $\delta$ signifies 
UV divergent counterterms, while with the symbol $\Delta$ 
we denote finite shifts.
A one-loop discussion 
of $\Delta v_a$ will be presented in 
section~\ref{renormalization}.
The shift leads to the scalar potential, including counterterms, 
\begin{subequations}\label{V}
\begin{eqnarray} 
V + \delta V - V_0 &=& 
\mc^{-\varepsilon/2} \left( t_a + \Delta t_a + \delta\mu^2_{ab} \bar v_b + 
\delta\lambda_{abcd}\, \bar v_b \bar v_c \bar v_d \right)  h_a
\\ && \label{V2}
+ \frac{1}{2} \left( \left( M^2_0 \right)_{ab} + 
\left( \Delta M^2_0 \right)_{ab} + \delta \mu^2_{ab} + 
3 \delta\lambda_{abcd}\, \bar v_c \bar v_d \right) h_a h_b
\\ &&
+ \mc^{\varepsilon/2} \left( \lambda_{abcd} + \delta\lambda_{abcd} \right)
\bar v_d  h_a h_b h_c 
\\ &&
+ \frac{1}{4} \mc^\varepsilon \left( \lambda_{abcd} + \delta\lambda_{abcd} \right)
h_a h_b h_c h_d,
\end{eqnarray}
\end{subequations}
with
$V$ as in equation~(\ref{V-tree}),
\begin{equation}\label{eq:tadpole-parameter-defs}
t_a = \mu^2_{ab} v_b + \lambda_{abcd} v_b v_c v_d, \quad
\Delta t_a = \mu^2_{ab} \bar v_b + \lambda_{abcd} \bar v_b \bar v_c \bar v_d 
- t_a, 
\end{equation}
$V_0$ being the constant term, 
\begin{equation}\label{M02}
\left( M^2_0 \right)_{ab} \equiv 
\mu^2_{ab} + 3 \lambda_{abcd} v_c v_d
\quad \mbox{and} \quad
\left( \Delta M^2_0 \right)_{ab} \equiv 
\mu^2_{ab} + 3 \lambda_{abcd} \bar v_c \bar v_d -
\left( M^2_0 \right)_{ab}.
\end{equation}
The quantities $\Delta t_a$ and $\left( \Delta M^2_0 \right)_{ab}$ 
will become useful when we go beyond the tree level because 
they will be induced by the shifts $\Delta v_a$.
We will drop $V_0$ in the rest of the paper since it does not alter
the dynamics of the theory.

From now on we choose the $v_a$ as the tree-level vacuum expectation 
values (VEVs) of the scalars,
\textit{i.e.}\ as the values of the $\varphi_a$ at the minimum of $V(\varphi)$. 
Taking the derivative of the scalar potential $V$, we obtain
\begin{equation}
\frac{\partial V}{\partial \varphi_a} = 
\mu^2_{ab} \varphi_b + \mc^\varepsilon \lambda_{abcd} \varphi_b \varphi_c \varphi_d.
\end{equation}
Therefore, the conditions that the $v_a$ ($a=1,\ldots,n_\varphi$) 
correspond to a stationary point of $V$ are given by 
\begin{equation}\label{eq:tadpole-condition}
t_a = 0 \quad \mbox{for} \quad a=1,\ldots,n_\varphi.
\end{equation}
SSB occurs if the minimum 
$\varphi_1 = v_1, \ldots, \varphi_{n_\varphi} = v_{n_\varphi}$
of $V$ is non-trivial, \textit{i.e.}\ 
different from $v_1 = \cdots = v_{n_\varphi} = 0$. 
In any case, whether there is SSB or not, $M^2_0$ of equation~(\ref{M02}) 
is the tree level mass matrix of the scalars.

The mass matrix of the fermions is given by
\begin{equation}
m_0 = \sum_{a=1}^{n_\varphi} v_a Y_a.
\end{equation}
The subscript $0$ in $m_0$ and $M^2_0$ indicates tree level 
mass matrices.
The tree-level mass matrices and fermions and scalars are diagonalized by
\begin{subequations}\label{diagonal}
\begin{eqnarray}
U_0^T m_0 U_0 &=& \hat{m}_0 \equiv 
\diag \left( m_{01}, \ldots, m_{0 n_\chi} \right), \\
W_0^T M^2_0 W_0 &=& {\hat M}^2_0 \equiv 
\diag \left( M^2_{01}, \ldots, M^2_{0 n_\varphi} \right),
\end{eqnarray}
\end{subequations}
where $U_0$ is unitary~\cite{Schur} and $W_0$ is orthogonal.

The diagonalization matrices $U_0$ and $W_0$ 
allow us to introduce mass eigenfields $\hat\chi_{jL}$ and $\hat h_a$
via 
\begin{equation}
\chi_{iL} = \left( U_0 \right)_{ij} \hat\chi_{jL} 
\quad \mbox{and} \quad
h_a = (W_0)_{ab} \hat h_b,
\end{equation}
respectively.
Rewriting the Lagrangian in terms of the mass eigenfields amounts to
the replacements
\begin{subequations}\label{eq:param-mass-basis}
\begin{align}
\delta^{(\chi)} &\rightarrow \hat{\delta}^{(\chi)} = 
U_0^\dagger \delta^{(\chi)} U_0, \label{eq:hat-delta-chi}\\
Y_a &\rightarrow \hat{Y}_a = \left(U_0^T  Y_{b} U_0 \right) (W_0)_{ba}, 
\label{eq:hat-Y}\\ 
\delta^{(\varphi)} &\rightarrow \hat{\delta}^{(\varphi)} = W_0^T
\delta^{(\varphi)} W_0,\\ 
v_a &\rightarrow \hat{v}_a = (W_0)_{ba} v_{b}, \label{vhat} \\
t_a &\rightarrow \hat{t}_a = (W_0)_{ba} t_{b},\\
\mu^2 &\rightarrow \hat{\mu}^2 = W_0^T \mu^2 W_0, \\ 
\lambda_{abcd} &\rightarrow \hat{\lambda}_{abcd}
= \lambda_{a'b'c'd'}(W_0)_{a'a} (W_0)_{b'b} (W_0)_{c'c} (W_0)_{d'd},
\end{align}
\end{subequations}
such that the form of the Lagrangian is preserved.
Therefore, without loss of generality we assume that we are 
in the mass bases of fermions and scalars, when we perform the 
one-loop computation of the selfenergies.
Note that $\hat{\bar{v}}_a$ and $\Delta \hat v_a$ are defined analogously to 
$\hat v_a$.

In the mass basis it is useful to rewrite the Yukawa interaction as
\begin{equation}\label{Y}
\mathcal{L}_Y = - \frac{1}{2} \bar{\hat\chi} 
\left( \hat Y_a \gamma_L + \hat Y_a^* \gamma_R \right) \hat\chi 
\left( \mc^{\varepsilon/2} \hat h_a + \hat{\bar{v}}_a \right)
\end{equation}
with
\begin{equation}\label{mef-m}
\gamma_L = \frac{\bone - \gamma_5}{2}, \quad
\gamma_R = \frac{\bone + \gamma_5}{2}, \quad
\hat\chi = \left( \begin{array}{c} 
\hat\chi_1 \\ \vdots \\ \hat\chi_{n_\chi} 
\end{array} \right)
\quad \mbox{and} \quad
\hat\chi_i = \hat\chi_{iL} +  \left( \hat\chi_{iL} \right)^c,
\end{equation}
where the superscript $c$ indicates charge conjugation.

\section{Renormalization}
\label{renormalization}
\paragraph{General outline:}
Our objective is to describe the general renormalization procedure and
to work out a prescription for the computation 
of the one-loop contribution to the \emph{physical} fermion and scalar masses.
For this purpose we have to compute the selfenergies. Clearly, 
the manner in which the selfenergies---and thus the quantities 
we aim at---depend on the parameters of our toy model is 
renormalization-scheme-dependent. It is, therefore, expedient to clearly 
expound the scheme we want to use and how we plan to reach our goal. 

We proceed in three steps:
\begin{enumerate}
\renewcommand{\theenumi}{\roman{enumi}}
\renewcommand{\labelenumi}{\roman{enumi}.}
\item \label{li:Ren-List-1}
$\overline{\mbox{MS}}$ renormalization for the determination of 
$\delta {\hat{Y}}_a$, $\delta{\hat{\lambda}}_{abcd}$, 
${\delta\hat{\mu}}^2_{ab}$, $\hat{\delta}^{(\chi)}$ and 
$\hat{\delta}^{(\varphi)}$.
\item \label{li:Ren-List-2}
Finite shifts $\Delta \hat{v}_a$ such that the scalar one-point 
functions of the $\hat{h}_a$ are zero. 
These two steps allow us to compute \emph{renormalized} 
one-loop selfenergies $\Sigma(p)$ and $\Pi(p^2)$
for fermions and scalars, respectively.
\item
Finite field strength renormalization in order to switch from the
$\overline{\mbox{MS}}$ selfenergies $\Sigma(p)$ and $\Pi(p^2)$ 
to on-shell selfenergies\footnote{Note that here the term \textit{on-shell} 
  refers to field strength renormalization only. 
  We have no mass counterterms, because in our approach masses are derived
  quantities and, therefore, functions of the parameters of the model---see
  the discussion at the end of this section.}
$\widetilde\Sigma(p)$ and $\widetilde\Pi(p^2)$.
\end{enumerate}

Several remarks are in order to concretize this outline.
$\overline{\mbox{MS}}$ renormalization, \textit{i.e.}\ subtraction of terms
proportional to the constant 
\begin{equation}
c_\infty = \frac{2}{\varepsilon} - \gamma + \ln(4\pi),
\end{equation}
where $\gamma$ is the Euler--Mascheroni constant,
is realized in the following way:
\begin{enumerate}
\renewcommand{\labelenumi}{(\alph{enumi})}
\item
$\delta{\hat{\lambda}}_{abcd}$ is determined from the quartic scalar coupling,
\item
$\delta {\hat{Y}}_a$ is obtained from the Yukawa vertex,
\item
$\delta {\hat{\mu}}^2_{ab}$ removes $c_\infty$ from the $p^2$-independent part
of the scalar selfenergy,
\item
$\hat{\delta}^{(\chi)}$ and $\hat{\delta}^{(\varphi)}$ are determined from the 
momentum-dependent parts of the respective selfenergies. 
\end{enumerate}
With the prescriptions~(a)--(d) above, all correlation functions and 
all physical quantities computed in our toy model must be finite. This applies
in particular to the selfenergies.

\paragraph{Fermion selfenergy:}
Let us first consider the renormalized fermion selfenergy $\Sigma(p)$,
defined via the inverse propagator matrix
\begin{equation}\label{inprop-fermion}
S^{-1}(p) = \slashed{p} - \hat{m}_0 - \Sigma(p),
\end{equation}
where $\Sigma(p)$ has the chiral structure 
\begin{equation}\label{sigma-fermion}
\Sigma(p) = \slashed{p} \left( \Sigma^{(A)}_L(p^2) \gamma_L + 
 \Sigma^{(A)}_R(p^2) \gamma_R \right) + 
\Sigma^{(B)}_L(p^2) \gamma_L + \Sigma^{(B)}_R(p^2) \gamma_R.
\end{equation}
For the relationships between $\Sigma^{(A)}_L$ and $\Sigma^{(A)}_R$ and 
between $\Sigma^{(B)}_L$ and $\Sigma^{(B)}_R$ in the case of Dirac and 
Majorana fermions we refer 
the reader to appendix~\ref{selfenergies}.
At one-loop order, $\Sigma(p)$ has the terms 
\begin{eqnarray}
\label{Sigma}
\Sigma(p) &=&
\Sigma^{\textrm{1-loop}}(p) 
- \slashed{p} \left[ \hat{\delta}^{(\chi)} \gamma_L + 
\left( \hat{\delta}^{(\chi)} \right)^* \gamma_R \right]
\nonumber \\* && 
+ \hat{v}_a \left[ \delta {\hat{Y}}_a \gamma_L + 
(\delta {\hat{Y}}_a)^* \gamma_R \right] 
+ \Delta \hat{v}_a \left[ {\hat{Y}}_a \gamma_L + {\hat{Y}}_a^* \gamma_R \right], 
\end{eqnarray}
where $\Sigma^{\textrm{1-loop}}$ corresponds to the diagram of 
figure~\ref{1loop-f}.
Since $\delta {\hat{Y}}_a$ is already determined by the Yukawa vertex, the
corresponding term in $\Sigma(p)$ must automatically make 
$\Sigma^{(B)}_{L,R}$ in equation~(\ref{Sigma}) finite. 
As for $\Sigma^{(A)}_{L,R}$ in $\Sigma(p)$, we note that these matrices 
are hermitian---see also appendix~\ref{selfenergies}, therefore,
the counterterms with the hermitian matrix $\hat{\delta}^{(\chi)}$ 
suffice for finiteness.
The last term in equation~(\ref{Sigma}) is induced by the finite VEV shifts.

\paragraph{Scalar selfenergy:}
Now we address the inverse scalar propagator matrix
\begin{equation}\label{invprop-scalar}
\Delta^{-1}(p^2) = p^2 - {\hat M}^2_0 - \Pi(p^2).
\end{equation}
The scalar selfenergy $\Pi(p^2)$ has the structure
\begin{equation}
\label{Pi}
\Pi_{ab}(p^2) = \Pi^{\textrm{1-loop}}_{ab}(p^2) - \hat{\delta}^{(\varphi)}_{ab} p^2 + 
\delta{\hat{\mu}}^2_{ab} + 3 \delta{\hat{\lambda}}_{abcd} \hat{v}_c \hat{v}_d + 
6 {\hat{\lambda}}_{abcd} \hat{v}_c \Delta \hat{v}_d
\end{equation}
at one-loop order. 
With an argument analogous to the fermionic case we find that the symmetric 
matrix $\hat{\delta}^{(\varphi)}$ suffices for making the derivative of $\Pi(p^2)$
finite. According to our renormalization
prescription, $\delta{\hat{\lambda}}_{abcd} \hat{v}_c \hat{v}_d$ 
is already fixed, but we have
$\delta{\hat{\mu}}^2_{ab}$ at our disposal to cancel the infinity in the
$p^2$-independent term in $\Pi(p^2)$.
The last term in the scalar selfenergy, equation~(\ref{Pi}), 
stems from the finite mass corrections 
$\Delta M^2_0$---see equation~(\ref{V2})---expressed 
in terms of the finite VEV shifts 
induced by tadpole contributions.

Another commonly used approach for the renormalization of $ {\hat{\mu}}^2$, 
\textit{e.g.}\ in~\cite{Krause:2016oke}, is to express its diagonal entries 
via the tadpole parameters $\hat{t}_a$ as of 
equation~(\ref{eq:tadpole-parameter-defs}), 
resulting in renormalization conditions more closely related to 
physical observables.
However, there are simply not enough tadpole parameters available to replace 
\emph{all} parameters in the $n_\varphi \times n_\varphi$ symmetric 
matrix ${\hat{\mu}}^2$ and we have 
two main reasons for dismissing this choice in our case.
One is that expressing $ {\hat{\mu}}^2_{aa}$ in terms of the tadpole parameters 
involves the inverses of the VEVs $\hat{v}_a$. 
In the general case, some of these can be zero, 
leading to ill-defined expressions for $\delta {\hat{\mu}}^2_{aa}$. 
The other one is that the diagonal and off-diagonal entries of 
${\hat{\mu}}^2$ can be treated on an equal footing in our approach, 
leading to a more compact description.

\paragraph{One-point function:}
These shifts derive from the linear term in the scalar potential.
For simplicity we stick to 
the lowest non-trivial order, where it is given by 
\begin{equation}\label{t1}
\mc^{-\varepsilon/2} \left( \hat{t}_a + \Delta \hat{t}_a + 
\delta{\hat{\mu}}^2_{ab} \hat{v}_b + 
\delta{\hat{\lambda}}_{abcd}\, \hat{v}_b \hat{v}_c \hat{v}_d \right) \hat h_a.
\end{equation}
Diagrammatically, the one-point function pertaining to $\hat h_a$ 
has the contributions\footnote{We stress again that we do not introduce 
  tadpole counterterms.} 
\begin{eqnarray} \label{eq:tadpole-condition-graph}
\lefteqn{
\begin{fmffile}{tadpoles-sum-general}
      \begin{gathered}
      \begin{fmfgraph*}(40,40)
	\fmftop{i1} 
	\fmfbottom{o1}
	\fmf{phantom}{i1,v1,o1}
	\fmf{dashes}{o1,v1}
	\fmf{phantom,left}{v1,i1,v1}
	\fmfv{decor.shape=circle,decor.size=3}{v1}
      \end{fmfgraph*}
    \end{gathered}
    +
      \begin{gathered}
      \begin{fmfgraph*}(40,40)
	\fmftop{i1} 
	\fmfbottom{o1}
	\fmf{phantom}{i1,v1,o1}
	\fmf{dashes}{o1,v1}
	\fmf{phantom,left}{v1,i1,v1}
	\fmfv{decor.shape=circle,decor.filled=shaded}{v1}
      \end{fmfgraph*}
    \end{gathered}
    + \begin{gathered}
      \begin{fmfgraph*}(40,40)
	\fmftop{i1}
	\fmfbottom{o1}
	\fmf{phantom}{i1,v1,o1}
	\fmf{dashes}{o1,v1}
	\fmf{phantom,left}{v1,i1,v1}
	\fmfv{decor.shape=cross}{v1}
      \end{fmfgraph*}
    \end{gathered} 
\end{fmffile}
} \nonumber
\\ &&
= \mc^{-\varepsilon/2}\frac{i}{-M^2_{0a}} \times
(-i)\left( \hat{t}_a + T_a + \Delta \hat{t}_a + 
\delta{\hat{\mu}}^2_{ab} \hat{v}_b + 
\delta{\hat{\lambda}}_{abcd}\, \hat{v}_b \hat{v}_c \hat{v}_d \right) = 0,
\end{eqnarray}
where $i/(-M^2_{oa})$ is the external scalar propagator at zero momentum.
The requirement that the one-point function is zero is identical with 
the requirement that the VEV of $\hat h_a$ is zero.
The first diagram in equation~(\ref{eq:tadpole-condition-graph})
represents the scalar tree-level 
one-point function corresponding to $\hat{t}_a$, 
which vanishes identically due to 
equation~(\ref{eq:tadpole-condition});
we have included it only for illustrative purposes.
The second diagram, which represents the one-loop 
tadpole contributions, corresponds to $T_a$. 
The third diagram represents the 
sum of $\Delta \hat t_a$ and the two counterterm contributions. 
We can decompose $T_a$ into 
an infinite and a finite part, \textit{i.e.}
\begin{equation}
T_a = \left( T_\infty \right)_a + \left( T_\mathrm{fin} \right)_a.
\end{equation}
Since with the imposition of
conditions~(a)--(d) the theory becomes finite, in equation~(\ref{t1}) 
we necessarily have 
\begin{equation}\label{T-finite}
\delta{\hat{\mu}}^2_{ab} \hat{v}_b + 
\delta{\hat{\lambda}}_{abcd}\, \hat{v}_b \hat{v}_c \hat{v}_d + 
\left( T_\infty \right)_a = 0.
\end{equation}
An explicit check of this relation is presented in section~\ref{cancellation}.
Moreover, we translate the finite tadpole contributions 
$\left( T_\mathrm{fin} \right)_a$ to shifts of the VEVs $\Delta \hat{v}_b$, 
similar to the approach of~\cite{Denner:2016etu}. 
At one-loop order this is effected by
\begin{equation}
\Delta \hat{t}_a = {\hat{\mu}}^2_{ab} \Delta \hat{v}_b + 3{\hat{\lambda}}_{abcd} \hat{v}_c \hat{v}_d \Delta \hat{v}_b = 
\left( {\hat M}^2_0 \right)_{ab} \Delta \hat{v}_b,
\end{equation}
where we have used equation~(\ref{eq:tadpole-parameter-defs}).
Therefore, equation~(\ref{eq:tadpole-condition-graph}) 
leads to the finite shift
\begin{equation}\label{eq:vev-shift}
\Delta \hat{v}_a = 
-\left( {\hat M}^2_0 \right)^{-1}_{ab} \left( T_\mathrm{fin} \right)_b.
\end{equation}
Note that these finite shifts eventually contribute to the 
finite mass corrections because they contribute to
the two-point functions of the fermions and scalars---see 
equations~(\ref{Sigma}) and~(\ref{Pi}), respectively.
Further clarifications concerning the VEV shifts $\Delta v_a$ 
are found in appendix~\ref{finite tadpole}.

\paragraph{Pole masses and finite field strength renormalization:}
It remains to perform a finite field strength renormalization in order to 
transform the one-loop selfenergies $\Sigma(p)$ and $\Pi(p^2)$ 
to on-shell selfenergies $\widetilde\Sigma(p)$ and $\widetilde\Pi(p^2)$, 
respectively. Immediately the question arises why we 
cannot use the $Z^{(1/2)}_\chi$ and $Z^{(1/2)}_\varphi$ defined in 
section~\ref{lagrangian} for this purpose. Note that 
we have incorporated these matrices into 
$\delta Y_a$ and $\delta \lambda_{abcd}$ at the respective interaction vertices. 
Therefore, in $\delta\mathcal{L}$ the field strength renormalization matrices 
$Z^{(1/2)}_\chi$ and $Z^{(1/2)}_\varphi$ occur solely in the hermitian matrix 
$\delta^{(\chi)}$ and the symmetric matrix $\delta^{(\varphi)}$, respectively.
Obviously, the latter matrices have fewer parameters than the original ones
and it is impossible to perform on-shell renormalization with
$\delta^{(\chi)}$ for more than one fermion field and with 
$\delta^{(\varphi)}$ for more than one scalar field.
What happens if we do not incorporate $Z^{(1/2)}_\chi$ and $Z^{(1/2)}_\varphi$ into 
the Yukawa and quartic couplings, respectively? Let us consider the Yukawa 
interaction for definiteness and denote by $\check\delta Y_a$ the Yukawa 
counterterm where $Z^{(1/2)}_\chi$ is not incorporated. Obviously, 
the relation between $\delta Y_a$ and $\check\delta Y_a$ is given by
\begin{equation}
\delta Y_a =
\left( Z_\chi^{(1/2)} \right)^T \left( Y_b + \check\delta Y_b \right) Z_\chi^{(1/2)} 
\left( Z^{(1/2)}_\varphi \right)_{ba} - Y_a.
\end{equation}
Actually, the quantity that is determined by the 
$\overline{\mbox{MS}}$ Yukawa vertex renormalization 
is $\delta Y_a$ and not $\check\delta Y_a$. 
Moreover, since we generate mass terms by SSB, 
the fermion mass term is induced by the shift of equation~(\ref{shift}) 
and has the form
\begin{equation}
\frac{1}{2} \chi_L^T C^{-1} \delta Y_a \bar{v}_a \chi_L + 
\frac{1}{2} \chi_L^T C^{-1} Y_a \bar v_a \chi_L + \mbox{H.c.}
\end{equation}
Thus it is clearly the same $\delta Y_a$ that occurs in both 
the mass term and the vertex renormalization. 
Consequently, with the counterterms of the 
unbroken theory we always end up with $\delta Y_a$ and $\delta^{(\chi)}$ 
as independent quantities and we 
can in general not perform on-shell renormalization.
Therefore, we need, in addition to $Z^{(1/2)}_\chi$ and $Z^{(1/2)}_\varphi$,
finite field strength renormalization matrices 
$\overset{\circ}{Z}^{\raisebox{-6pt}{$\scriptstyle (1/2)$}}_\chi$
and 
$\overset{\circ}{Z}^{\raisebox{-6pt}{$\scriptstyle (1/2)$}}_h$
for fermions and bosons, respectively,
inserted into the \emph{broken} Lagrangian, in order to perform on-shell 
renormalization.
In this way, the $\sqrt{Z}$-factors of the external lines in 
the LSZ formalism are exactly one~\cite{Aoki:1982}. 

We denote the one-loop contributions to
$\overset{\circ}{Z}^{\raisebox{-6pt}{$\scriptstyle (1/2)$}}_\chi$
and 
$\overset{\circ}{Z}^{\raisebox{-6pt}{$\scriptstyle (1/2)$}}_h$
by
$\frac{1}{2}\overset{\circ}{z}_\chi$ and 
$\frac{1}{2}\overset{\circ}{z}_h $, respectively. 
For the details of the computation and the results for 
these quantities we refer the reader to appendix~\ref{selfenergies}.
Here we only state the masses~\cite{kiyoura,Grimus:2016hmw}
\begin{eqnarray}\label{eq:one-loop-masses}
m_i &=& m_{0i} + m_{0i} \left( \Sigma^{(A)}_L \right)_{ii}(m^2_{0i}) +
\mbox{Re} \left( \Sigma^{(B)}_L \right)_{ii}(m^2_{0i}),
\\
M^2_a &=& M^2_{0a} + \Pi_{aa}(M^2_{0a})
\end{eqnarray}
at one-loop order. 
There is no summation in these two formulas over equal indices. 

Eventually we remark that one could decompose $\overset{\circ}{z}_\chi$ into 
a hermitian and an antihermitian matrix. One could be tempted to conceive
the antihermitian part as a correction to the tree-level diagonalization 
matrix $U_0$. However, we think that in our simple model such a decomposition 
has no physical meaning; in essence, we have no PMNS mixing matrix at disposal 
where it could become physical.
Of course, a similar remark applies to $\overset{\circ}{z}_h$---see 
also~\cite{altenkamp} for a recent discussion in the context of 
the two-Higgs-doublet model.

\section{Renormalization at the one-loop level}
\label{one-loop}
In this section we concretize, at the one-loop level, 
the renormalization procedure introduced in the previous section.  
For the relevant integrals needed for these computations 
see appendix~\ref{integrals}.

\subsection{One-loop results for selfenergies and tadpoles}
\begin{figure}[ht]
\begin{center}
  \begin{fmffile}{fermion-selfenergy}
      \begin{fmfgraph*}(100,60)
        \fmfstraight
	\fmfbottom{i,x1,x2,o}
	\fmf{plain,tension=5}{i,v1}
	\fmf{plain,tension=5}{v2,o}
	\fmf{plain}{v1,v2}
	\fmf{dashes,left}{v1,v2}
      \end{fmfgraph*}
  \end{fmffile}
\caption{One-loop fermion selfenergy diagram.}
\label{1loop-f}
\end{center}
\end{figure}
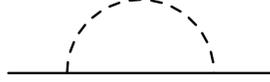

Here we display the results for the one-loop contributions 
$\Sigma^{\textrm{1-loop}}(p)$ and  $\Pi^{\textrm{1-loop}}_{ab}(p^2)$ 
to the fermion and scalar selfenergies, respectively, and also for the one-loop 
tadpole expression $T_a$.
\paragraph{Fermion selfenergy:}
The only direct one-loop contribution to the fermionic self-energy 
is given by the diagram of figure~\ref{1loop-f}.
Then, the definitions 
\begin{equation}
\Delta_{a,k} = x M^2_{0a} + (1-x) m^2_{0k} - x(1-x) p^2,
\end{equation}
\begin{equation}
D_{a,k} = \int_0^1 \mathrm{d}x\, x \ln \frac{\Delta_{a,k}}{\mc^2}, \quad
E_{a,k} = \int_0^1 \mathrm{d}x \ln \frac{\Delta_{a,k}}{\mc^2}
\end{equation}
and 
\begin{equation}
\hat D_a = \diag \left( D_{a,1}, \ldots, D_{a,n_\chi} \right), \quad
\hat E_a = \diag \left( E_{a,1}, \ldots, E_{a,n_\chi} \right)
\end{equation}
allow us to write the one-loop contribution to the fermionic selfenergy as
\begin{subequations}\label{sigma-1-loop}
\begin{eqnarray}
\Sigma^{\textrm{1-loop}} &=&
\frac{1}{16 \pi^2} \left\{
\slashed{p} \gamma_L \left[ -\frac{1}{2} c_\infty {\hat{Y}}_a^* {\hat{Y}}_a 
+ {\hat{Y}}_a^* \hat D_a {\hat{Y}}_a \right] \right.
\\ &&
+ \slashed{p} \gamma_R \left[ -\frac{1}{2} c_\infty {\hat{Y}}_a {\hat{Y}}_a^* 
+ {\hat{Y}}_a \hat D_a {\hat{Y}}_a^* \right]
\label{delta-chi} \\ &&
+ \gamma_L \left[ -c_\infty {\hat{Y}}_a \hat{m}_0 {\hat{Y}}_a + 
{\hat{Y}}_a \hat{m}_0 \hat E_a {\hat{Y}}_a \right] 
\label{div-sigma-B} \\ &&
\left. + \gamma_R \left[ -c_\infty {\hat{Y}}_a^* \hat{m}_0 {\hat{Y}}_a^* + 
{\hat{Y}}_a^* \hat{m}_0 \hat E_a {\hat{Y}}_a^* \right] \right\}.
\end{eqnarray}
\end{subequations}
\paragraph{Scalar selfenergy:}
 \begin{figure}[t]
  \begin{fmffile}{scalar-selfenergy}
  \begin{center}
  \begin{subfigure}[t]{.3\textwidth}\centering
      \begin{fmfgraph*}(100,60)
	\fmfleft{i}
	\fmfright{o}
	\fmf{phantom}{i,v1,v3,v4,v2,o}
	\fmf{dashes}{i,v1}
	\fmf{dashes}{v2,o}
	\fmf{plain,left}{v1,v2}
	\fmf{plain,left}{v2,v1}
      \end{fmfgraph*}
    \caption{}
    \end{subfigure}
    \begin{subfigure}[t]{.3\textwidth} \centering
       \begin{fmfgraph*}(100,60)
	\fmfleft{i}
	\fmfright{o}
	\fmf{phantom}{i,v1,v3,v4,v2,o}
	\fmf{dashes}{i,v1}
	\fmf{dashes}{v2,o}
	\fmf{dashes,left}{v1,v2}
	\fmf{dashes,left}{v2,v1}
      \end{fmfgraph*}
      \caption{}
    \end{subfigure}
    \begin{subfigure}[t]{.3\textwidth} \centering
      \begin{fmfgraph*}(100,60)
	\fmfleft{i}
	\fmfright{o}
	\fmftop{t}
	\fmf{dashes}{i,v,o}
	\fmffreeze
	\fmf{dashes,left=.9}{v,t,v}
	\fmfv{decor.shape=circle,decor.size=2}{v}
      \end{fmfgraph*}
      \caption{}
    \end{subfigure}
  \end{center}
  \end{fmffile}
  \caption{The Feynman diagrams of the one-loop contributions to the scalar selfenergy.}
  \label{scalar-selfenergy-graphs}
\end{figure}
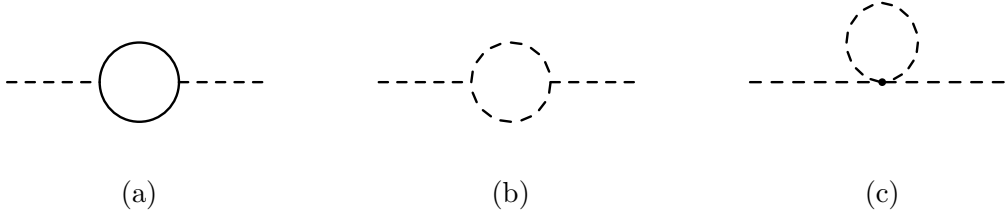

In the following, the superscripts $(a)$, $(b)$, $(c)$ 
refer to the Feynman diagrams of figure~\ref{scalar-selfenergy-graphs}. 
Thus the selfenergy has the contributions
\begin{equation}
\Pi^{\textrm{1-loop}}_{ab}(p^2) = 
\Pi^{(a)}_{ab}(p^2) + \Pi^{(b)}_{ab}(p^2) + \Pi^{(c)}_{ab}(p^2).
\end{equation}
We define 
\begin{equation}
\Delta_{ij} = x m_{0i}^2 + (1-x) m_{0j}^2 - x(1-x) p^2
\quad \mbox{and} \quad
\tilde\Delta_{rs} = x M_{0r}^2 + (1-x) M_{0s}^2 - x(1-x) p^2.
\end{equation}
With these definitions we obtain
\begin{subequations}
\begin{eqnarray}
\Pi^{(a)}_{ab}(p^2) &=&
\frac{1}{16 \pi^2} \left\{ c_\infty \,\mbox{Tr} \left[
{\hat{Y}}_a {\hat{m}_0} {\hat{Y}}_b {\hat{m}_0} + 
{\hat{Y}}_a^* \hat{m}_0 {\hat{Y}}_b^* \hat{m}_0 + 
2 {\hat{Y}}_a {\hat{Y}}_b^* {\hat{m}_0}^2  + 
2 {\hat{Y}}_a^* {\hat{Y}}_b {\hat{m}_0}^2 \right] 
\vphantom{\ln \frac{\Delta_{ij}}{\mc^2}}\right.
\nonumber \\ &&
-\frac{1}{2} c_\infty \, \mbox{Tr} \left[ {\hat{Y}}_a {\hat{Y}}_b^* + 
{\hat{Y}}_a^* {\hat{Y}}_b \right] p^2
\label{delta-h} \nonumber \\ &&
+\mbox{Tr} \left[ \left( {\hat{Y}}_a {\hat{Y}}_b^* + 
{\hat{Y}}_a^* {\hat{Y}}_b \right)  
\left( {\hat{m}_0}^2 - \frac{1}{6}\, p^2 \right)\right]
\nonumber \\ &&
- \int_0^1 \mathrm{d}x 
\left[ \left( ({\hat{Y}}_a)_{ij} ({\hat{Y}}_b)_{ji}^* + 
({\hat{Y}}_a)_{ij}^* ({\hat{Y}}_b)_{ji} \right)
\left( 2 \Delta_{ij} - x(1-x) p^2 \right) \right.
\nonumber \\* &&
\left. \left. + ({\hat{Y}}_a)_{ij} m_{0j} ({\hat{Y}}_b)_{ji} m_{0i} + 
({\hat{Y}}_a)_{ij}^* m_{0j} ({\hat{Y}}_b)_{ji}^* m_{0i} \right] 
\ln \frac{\Delta_{ij}}{\mc^2} \right\}, \label{Pi-a}
\\
\Pi^{(b)}_{ab}(p^2) &=& 
-\frac{18}{16 \pi^2} \,{\hat{\lambda}}_{acrs} \hat{v}_c {\hat{\lambda}}_{bdrs} 
\hat{v}_d
\left( c_\infty - \int_0^1 \mathrm{d}x\,\ln\frac{\tilde\Delta_{rs}}{\mc^2} 
\right),\label{Pi-b}
\\
\Pi^{(c)}_{ab}(p^2) &=&
-\frac{3}{16 \pi^2}\,{\hat{\lambda}}_{abrr} 
{M}^2_{0r} \left( c_\infty + 1 - 
\ln \frac{M^2_{0r}}{\mc^2} \right). \label{Pi-c}
\end{eqnarray}
\end{subequations}
For the following discussion, it is useful to introduce a separate notation 
for the divergent $p^2$-independent parts of $\Pi^{\textrm{1-loop}}_{ab}$:
\begin{subequations}
\begin{eqnarray}
\left( \Pi^{(a)}_\infty \right)_{ab}  &=&
\frac{1}{16 \pi^2} c_\infty \,\mbox{Tr} \left[
{\hat{Y}}_a \hat{m}_0 {\hat{Y}}_b \hat{m}_0 + 
{\hat{Y}}_a^* \hat{m}_0 {\hat{Y}}_b^* \hat{m}_0 + 
2 {\hat{Y}}_a {\hat{Y}}_b^* {\hat{m}_0}^2  + 
2 {\hat{Y}}_a^* {\hat{Y}}_b {\hat{m}_0}^2 \right],  
\label{div-a} \\
\left( \Pi^{(b)}_\infty \right)_{ab}  &=&
-\frac{18}{16 \pi^2} c_\infty \,
{\hat{\lambda}}_{acrs} \hat{v}_c {\hat{\lambda}}_{bdrs} \hat{v}_d,
\label{div-b} \\
\left( \Pi^{(c)}_\infty \right)_{ab}  &=&
-\frac{3}{16 \pi^2} c_\infty \,{\hat{\lambda}}_{abrr} M^2_{0r}.
\label{div-c}
\end{eqnarray}
\end{subequations}

\paragraph{Tadpoles:}
There are two one-loop tadpole contributions to the scalar one-point function,
namely
\begin{equation} \label{eq:tadpole-graphs}
\begin{fmffile}{tadpoles-graphs}
      \begin{gathered}
      \begin{fmfgraph*}(40,40)
	\fmftop{i1} 
	\fmfbottom{o1}
	\fmf{phantom}{i1,v1,o1}
	\fmf{dashes}{o1,v1}
	\fmf{plain,left}{v1,i1,v1}
      \end{fmfgraph*}
    \end{gathered}
    + \begin{gathered}
      \begin{fmfgraph*}(40,40)
	\fmftop{i1} 
	\fmfbottom{o1}
	\fmf{phantom}{i1,v1,o1}
	\fmf{dashes}{o1,v1}
	\fmf{dashes,left}{v1,i1,v1}
      \end{fmfgraph*}
    \end{gathered}
    = \mc^{-\varepsilon/2} \frac{i}{-M_{0a}^2} \times (-i) 
\left( T^{(\chi)}_a + T^{(h)}_a \right).
\end{fmffile}
\end{equation}
We find the following result for tadpole terms:
\begin{eqnarray}
T^{(\chi)}_a &=& \frac{1}{16 \pi^2}\, \mbox{Tr} 
\left[ \left( {\hat{Y}}_a {\hat{m}^3_0} + {\hat{Y}}_a^*{\hat{m}^3_0} \right) 
\left( c_\infty + 1 - \ln \frac{{\hat{m}^2_0}}{\mc^2} \right) \right],
\label{Tchi} \\
T^{(h)}_a &=& -\frac{3}{16 \pi^2} {\hat{\lambda}}_{abrr} \hat{v}_b {M}^2_{0r}
\left( c_\infty + 1 - \ln \frac{M^2_{0r}}{\mc^2} \right).
\label{Th}
\end{eqnarray}
We denote the divergences in the tadpole expressions by
$\left( T^{(\chi)}_\infty \right)_a$ and $\left( T^{(h)}_\infty \right)_a$.

\subsection{Determination of the counterterms}\label{determination}
\paragraph{Counterterms of Yukawa and quartic scalar couplings:}
Using $\overline{\mathrm{MS}}$ renormalization, it is straightforward to 
compute these counterterms.
For the Yukawa couplings we obtain 
\begin{equation}\label{dy}
\delta {\hat{Y}}_a = 
\frac{1}{16 \pi^2} c_\infty {\hat{Y}}_b {\hat{Y}}_a^* {\hat{Y}}_b.
\end{equation}
The $\delta{\hat{\lambda}}_{abcd}$ can be split into
\begin{equation}\label{dl}
\delta{\hat{\lambda}}_{abcd} = 
\delta{\hat{\lambda}}^{(\chi)}_{abcd} + \delta{\hat{\lambda}}^{(\varphi)}_{abcd},
\end{equation}
generated by fermions and scalars, respectively, in the loop.
The first case yields 
\begin{eqnarray}
\delta{\hat{\lambda}}^{(\chi)}_{abcd} &=& -\frac{1}{3} \times
\frac{1}{16 \pi^2} c_\infty 
\mathrm{Tr} \left[ 
{\hat{Y}}_a {\hat{Y}}_b^* {\hat{Y}}_c {\hat{Y}}_d^* + 
{\hat{Y}}_a {\hat{Y}}_c^* {\hat{Y}}_d {\hat{Y}}_b^* + 
{\hat{Y}}_a {\hat{Y}}_d^* {\hat{Y}}_b {\hat{Y}}_c^* \right. \nonumber \\
&& \left. + 
{\hat{Y}}_a^* {\hat{Y}}_b {\hat{Y}}_c^* {\hat{Y}}_d + 
{\hat{Y}}_a^* {\hat{Y}}_c {\hat{Y}}_d^* {\hat{Y}}_b + 
{\hat{Y}}_a^* {\hat{Y}}_d {\hat{Y}}_b^* {\hat{Y}}_c
\right].
\label{dlch}
\end{eqnarray}
In this formula we have taken into account that the Yukawa coupling matrices 
are symmetric. The scalar contribution is
\begin{equation}\label{dlph}
\delta{\hat{\lambda}}^{(\varphi)}_{abcd} = 
\frac{3}{16 \pi^2} c_\infty \left(
{\hat{\lambda}}_{abrs} {\hat{\lambda}}_{rscd}  + 
{\hat{\lambda}}_{adrs} {\hat{\lambda}}_{rsbc} + 
{\hat{\lambda}}_{acrs} {\hat{\lambda}}_{rsbd}\right).
\end{equation}
\paragraph{Counterterms pertaining to field strength renormalization:} 
Cancellation of the divergence in equation~(\ref{delta-chi}) 
determines $\hat{\delta}^{(\chi)}$ as 
\begin{equation}\label{c-delta-chi}
\hat{\delta}^{(\chi)} = 
-\frac{1}{2} \times \frac{1}{16 \pi^2} c_\infty {\hat{Y}}_a^* {\hat{Y}}_a.
\end{equation}
Considering the scalar selfenergy, we find that only diagram~(a) of 
figure~\ref{scalar-selfenergy-graphs} has a divergence proportional to $p^2$.
Therefore, we obtain from equation~(\ref{Pi-a}) 
\begin{equation}\label{c-delta-h}
 \hat{\delta}^{(\varphi)}_{ab} = -\frac{1}{2} \times \frac{1}{16\pi^2} c_\infty 
\mbox{Tr}\left[{\hat{Y}}_a {\hat{Y}}_b^* + {\hat{Y}}_a^* {\hat{Y}}_b\right].
\end{equation}
\paragraph{Counterterm pertaining to ${\hat{\mu}}^2_{ab}$:}
The counterterm $\delta{\hat{\mu}}^2_{ab}$ has to be determined by the
cancellations of the divergences of equations~(\ref{div-b}) and~(\ref{div-c}).
Thus we demand 
\begin{eqnarray}
0 &=& \lefteqn{\delta{\hat{\mu}}^2_{ab} + 
3 \delta{\hat{\lambda}}^{(\varphi)}_{abcd} \hat{v}_c \hat{v}_d +
\left( \Pi^{(b)}_\infty \right)_{ab} + \left( \Pi^{(c)}_\infty \right)_{ab}}
\nonumber \\ &=& 
\delta{\hat{\mu}}^2_{ab} + \frac{3}{16 \pi^2} c_\infty \left[
3 {\hat{\lambda}}_{abrs} {\hat{\lambda}}_{rscd} \hat{v}_c \hat{v}_d - 
{\hat{\lambda}}_{abrr} M^2_{0r} \right] 
\nonumber \\ &=&
\delta{\hat{\mu}}^2_{ab} + \frac{3}{16 \pi^2} c_\infty \left[ 
{\hat{\lambda}}_{abrs} \left( {\hat{\mu}}^2_{rs} + 
3 {\hat{\lambda}}_{rscd} \hat{v}_c \hat{v}_d - {\hat{\mu}}^2_{rs} \right) - 
{\hat{\lambda}}_{abrr} M^2_{0r} \right] 
\nonumber \\ &=&
\delta{\hat{\mu}}^2_{ab} - 
\frac{3}{16 \pi^2} c_\infty \, {\hat{\lambda}}_{abrs} {\hat{\mu}}^2_{rs}.
\end{eqnarray}
Therefore, $\delta{\hat{\mu}}^2_{ab}$ is fixed as 
\begin{equation}\label{eq:delta-mu}
\delta {\hat{\mu}}^2_{ab} = 
\frac{3}{16 \pi^2} c_\infty \, {\hat{\lambda}}_{abrs} {\hat{\mu}}^2_{rs}.
\end{equation}

\subsection{Cancellation of divergences}
\label{cancellation}
Having fixed all available counterterms, the remaining UV divergences in 
the selfenergies and tadpoles have to drop out. This is what 
we want to show in this subsection. 
\paragraph{Fermion selfenergy:}
With $\delta {\hat{Y}}_a$ of equation~(\ref{dy}) and 
\begin{equation}
\hat{v}_a \delta {\hat{Y}}_a = \frac{1}{16 \pi^2}
c_\infty {\hat{Y}}_b \hat{m}_0 {\hat{Y}}_b,
\end{equation}
we find that $\Sigma$ is finite without any mass renormalization, 
as it has to be. 
\paragraph{Scalar selfenergy:}
We have already treated the divergences~(\ref{div-b}) 
and~(\ref{div-c}), but there is still the divergence of equation~(\ref{div-a}).
However, it is easy to see that its cancellation in the selfenergy~(\ref{Pi})
is simply effected by 
\begin{equation}
\left( \Pi^{(a)}_\infty \right)_{ab} + 
3 \delta{\hat{\lambda}}^{(\chi)}_{abcd} \hat{v}_c \hat{v}_d = 0.
\end{equation}
\paragraph{Tadpoles:}
It remains to verify equation~(\ref{T-finite}). First we consider the result 
of the fermionic tadpole in equation~(\ref{Tchi}). 
Contracting the counterterm $\delta {\hat{\lambda}}^{(\chi)}_{abcd}$ 
of equation~(\ref{dlch}) with the VEVs and adding to it  
$\big( T^{(\chi)}_\infty \big)_a$ yields
\begin{equation}
\left( T^{(\chi)}_\infty \right)_a + 
\delta{\hat{\lambda}}^{(\chi)}_{abcd} \hat{v}_b \hat{v}_c \hat{v}_d = 
\left( T^{(\chi)}_\infty \right)_a - 
\frac{1}{16 \pi^2} c_\infty \, \mbox{Tr} 
\left[ \left( {\hat{Y}}_a \hat{m}_0^3 + {\hat{Y}}_a^* \hat{m}_0^3 \right)\right] 
= 0. 
\end{equation}
Similarly, 
using equations~(\ref{dlph}) and~(\ref{eq:delta-mu}),
the scalar tadpole contribution of equation~(\ref{Th}) 
is found to be finite via
\begin{eqnarray}
\lefteqn{
\left( T^{(h)}_\infty \right)_a + \delta {\hat{\mu}}^2_{ab} \hat{v}_b + 
\delta{\hat{\lambda}}^{(\varphi)}_{abcd} \hat{v}_b \hat{v}_c \hat{v}_d 
}
\nonumber \\ &=&
\left( T^{(h)}_\infty \right)_a +
\frac{3}{16\pi^2} c_\infty \left(  {\hat{\lambda}}_{abrs} 
{\hat{\mu}}^2_{rs} \hat{v}_b + 
3{\hat{\lambda}}_{abrs}{\hat{\lambda}}_{rscd} 
\hat{v}_b \hat{v}_c \hat{v}_d \right) 
\nonumber \\ 
&=& \left( T^{(h)}_\infty \right)_a +
\frac{3}{16\pi^2} c_\infty {\hat{\lambda}}_{abrr}\hat{v}_b {M}^2_{0r}  
\,=\, 0.
\end{eqnarray}

\subsection{Counterterms and UV divergences in a general basis}
The results for the selfenergies and counterterms shown 
in the previous sections are given in the mass bases. However, 
for a check of the cancellation of divergences 
it might be advantageous to have 
the divergences in a general basis. Such expressions can be obtained 
by using the parameter transformations~(\ref{eq:param-mass-basis}).

As an example, let us do this transformation in the case of
${\hat\delta}^{(\chi)}$ of equation~(\ref{c-delta-chi}), 
where one has to apply
\begin{align}
  \delta^{(\chi)} &= U_0 \hat{\delta}^{(\chi)} U_0^\dagger 
\nonumber \\
  &= -\frac{1}{2} \times 
\frac{1}{16 \pi^2} c_\infty U_0 \hat{Y}_a^* \hat{Y}_a U_0^\dagger 
\nonumber \\
  &= -\frac{1}{2} \times \frac{1}{16 \pi^2} c_\infty U_0 
\left( U_0^\dagger Y_b^* U_0^* \left(W_0\right)_{ba} \right) 
\left( \vphantom{U_0^\dagger}U_0^T Y_c U_0 \left(W_0\right)_{ca} \right) U_0^\dagger 
\nonumber \\
  &= -\frac{1}{2} \times \frac{1}{16 \pi^2} c_\infty Y_a^* Y_a.
\end{align}
In the case of the divergence in $\Sigma^{(B)}_L$---see 
equation~(\ref{div-sigma-B}), we
have to use the slightly different transformation
\begin{equation}
U_0^* \hat Y_a \hat m_0 \hat Y_a U_0^\dagger = Y_a m_0^* Y_a
\end{equation}
This explains that we have to be careful when a fermion mass term occurs
because in general
\begin{equation}
 v_a Y_a^* = m_0^* \neq v_a Y_a = m_0.
\end{equation}
This complication only arises in 
\begin{equation}\label{c1}
\left( T^{(\chi)}_\infty \right)_a = \frac{1}{16 \pi^2} c_\infty \, \mbox{Tr} 
\left[ {Y}_a {{m}_0}^*{{m}_0}{{m}_0}^* + 
{Y}_a^*{{m}_0}{{m}_0}^*{{m}_0}\right]
\end{equation}
and
\begin{equation}\label{c2}
\left( \Pi^{(a)}_\infty \right)_{ab}  =
\frac{1}{16 \pi^2} c_\infty \,\mbox{Tr} \left[
{Y}_a {{m}_0}^* {Y}_b {{m}}_0^* + {Y}_a^* {m}_0 {Y}_b^* {m}_0 + 
2 {Y}_a {Y}_b^* {m}_0 {{m}_0}^*  + 2 {Y}_a^* {Y}_b {{m}_0}^* {m}_0 \right].
\end{equation}
The divergences 
$\big(T^{(h)}_\infty\big)_a$, 
$\big( \Pi^{(b)}_\infty \big)_{ab}$ and 
$\big( \Pi^{(c)}_\infty \big)_{ab}$ 
are obtained in a general basis by simply removing the hats from all 
quantities and the same is true for all counterterms. 

\section{An example of a flavour symmetry}
\label{example}
Motivated by flavour models of the lepton sector~\cite{reviews}, we
will now consider a Lagrangian with a simple flavour symmetry and 
study how renormalization is affected in this case.
\subsection{Symmetry group and Lagrangian}
We assume the same number of Majorana and scalar fields, \textit{i.e.}\ 
$n_\chi = n_\varphi \equiv n$. 
In addition, we require $n \geq 2$.
Instead of the $\mathbbm{Z}_4$ symmetry of equation~(\ref{S}), which acts at
the same time on all fields, we will now postulate a $\mathbbm{Z}_4$ symmetry
for every index $a = 1, \ldots, n$:
\begin{equation}
\clam{\mathbb{Z}_{4}}_{a} : \quad 
\chi^\bb_{aL}\rightarrow i \chi^\bb_{aL}, \quad 
\varphi^\bb_{a} \rightarrow -\varphi^\bb_{a}, \quad
\chi^\bb_{bL}\rightarrow \chi^\bb_{bL}, \quad 
\varphi^\bb_{b} \rightarrow \varphi^\bb_{b} \; \forall\, b \neq a.
\end{equation}
This has the consequence that scalar fields with the same index occur in pairs
in the scalar potential. Note that now it is reasonable to use 
the same indices for both fermions and scalars. 
In addition, we assume that the Lagrangian is invariant
under simultaneous permutations of fermion and scalar fields.
Therefore, group-theoretically the symmetry group of the Lagrangian can be
conceived as
\begin{equation}
G_n = \clam{\mathbb{Z}_{4}}^n \rtimes S_n.
\end{equation}

With this flavour group, the bare Lagrangian has the form 
\begin{equation}
\mathcal{L}_B = \sum_{a=1}^n \left[
i\overline{\chi}_{aL}^\bb \slashed \partial \chi_{aL}^\bb + 
\frac{1}{2} \partial_{\mu}\varphi_{a}^\bb \partial^{\mu}\varphi_{a}^\bb + 
\frac{1}{2}\,y^\bb \clam{\chi_{aL}^\bb C^{-1} \chi_{aL}^\bb \varphi^\bb_{a} + 
\text{H.c.}} \right] - V_B,
\end{equation}
where the bare scalar potential can be written as
\begin{equation}
V_B = \frac{1}{2}\,\mu^{2} 
\sum_{a=1}^n \left( \varphi_{a}^\bb \right)^2 +
\frac{1}{4}\,\lambda 
\left( \sum_{a=1}^n \left( \varphi_{a}^\bb\right)^2 \right)^{2} + 
\frac{1}{4}\lambda' 
\sum_{a,b = 1}^n
\clam{\varphi_{a}^\bb}^{2} \clam{\varphi_{b}^{\bb}}^{2} 
\left(1-\delta_{ab}\right),
\end{equation}
where $\delta_{ab}$ is the Kronecker delta.

\subsection{Relation to the general model}
Due to the symmetry group $G_n$, we only have one Yukawa 
coupling constant and two quartic couplings. In order to use the general 
one-loop results, we have to establish the relation between 
the general model of section~(\ref{lagrangian}) and the
present example. For simplicity we now drop the superscript $(B)$ and keep in
mind that the following list applies not only to the renormalized coupling 
constants but also to the counterterms and the bare coupling constants:
\begin{subequations}\label{eq:relation-to-general}
  \begin{align}
  \clam{Y_a}_{bc} &= y\, \delta_{ab} \delta_{ac} \quad \forall a, \\
  \left( \mu^2 \right)_{ab} &= \mu^2 \delta_{ab}, \\[-2mm]
  \lambda_{aaaa} &= \lambda \quad \forall a \quad \text{and} \quad 
  \lambda_{aabb} = \frac{1}{3} 
\left( \lambda + \lambda' \right) \quad \forall a \neq b. 
\label{eq:relation-lambda}
  \end{align}
\end{subequations}
Note that now we just have one mass parameter $\mu^2$. Moreover,  
quartic couplings $\lambda_{abbb}$ with $a \neq b$ and those with three or
four different indices are zero. 
Without loss of generality we assume $y > 0$. 
In addition, we have to consider equation~(\ref{delta}), which now reads
\begin{equation}
\delta^{(\chi)}_{ab} = \delta^{(\chi)} \delta_{ab}, \quad
\delta^{(\varphi)}_{ab} = \delta^{(\varphi)} \delta_{ab}, 
\end{equation}
because due to the symmetry group $G_n$ only one field strength 
renormalization constant is allowed for each type of fields.

The results of section~\ref{one-loop}, found for the general Yukawa model, 
can directly be used for the present case by  
applying equation~(\ref{eq:relation-to-general}). In this way we obtain 
the counterterms
\begin{subequations}
\begin{eqnarray}
\delta y &=& 
\frac{1}{16\pi^{2}} c_\infty y^{3}, 
\\
\delta\lambda^{(\chi)} &=& -\frac{2}{16\pi^{2}}c_\infty y^{4}, 
\\
\delta\lambda^{(\varphi)} &=& 
\frac{1}{16\pi^{2}}c_\infty 
\left[ 9\lambda^{2} + \left( n-1 \right) \clam{\lambda+\lambda'}^{2} \right], 
\\
\left( \delta\lambda +\delta\lambda' \right)^{(\chi)} &=& 0, 
\\
\left( \delta\lambda +\delta\lambda' \right)^{(\varphi)} &=& 
\frac{1}{16\pi^{2}}c_\infty \left[   6\lambda\left(\lambda + 
\lambda'\right) +(n+2)(\lambda+\lambda')^{2} \right], 
\\
\delta\mu^{2} &=& \frac{\mu^2}{16\pi^{2}}c_\infty 
\left[  3 \lambda + (n-1) (\lambda+\lambda') \right], 
\end{eqnarray}
\end{subequations}
where the superscripts $(\chi)$ and $(\varphi)$ indicate fermions and scalars 
in the loop, respectively, in analogy to the notation in 
section~\ref{determination}. Field strength renormalization yields 
\begin{equation}
\delta^{(\chi)} = -\frac{1}{2} \times \frac{1}{16\pi^{2}} c_\infty y^2
\quad \mbox{and} \quad
\delta^{(\varphi)} = -\frac{1}{16\pi^{2}} c_\infty y^2.
\end{equation}

\subsection{Spontaneous symmetry breaking}
In order to have SSB we assume $\mu^2 < 0$.
For the vacuum expectation values we introduce the notation
\begin{equation}
v^{2} = \sum_{a=1}^n v_{a}^2.
\end{equation}
Obviously, for the scalar potential to be bounded from below we must have 
$\lambda > 0$, but $\lambda'$ can be positive or negative.

\paragraph{Case $\lambda' > 0$:}
Here, the minimum of the scalar potential is achieved when only one VEV is 
nonzero. Without loss of generality we assume
\begin{equation}
v_1 = v, \quad v_2 = \cdots = v_n = 0 \quad \Rightarrow \quad
v^2 = -\frac{\mu^2}{\lambda}.
\end{equation}
The symmetry breaking can be formulated as
\begin{equation}
G_n \xrightarrow{\text{SSB}} G_{n-1},
\end{equation}
where $G_{n-1}$ is the residual symmetry group. This residual symmetry is 
reflected in the mass spectrum
\begin{equation}\label{m+}
M^2_{01} = 2 \lambda v^2, \;
M^2_{02} = \cdots M^2_{0n} = \lambda' v^2, \quad
m_{01} = yv, \; m_{02} = \cdots = m_{0n} = 0.
\end{equation}

Since the mass matrices of both fermions and scalars are diagonal at 
tree level, it is straightforward to compute the one-loop corrections 
to equation~(\ref{m+}). 
It easy to see that at one-loop order the VEV shifts fulfill 
$\Delta v_2 = \cdots = \Delta v_n = 0$, only 
$\Delta v_1$ will in general be
nonzero. It is also obvious that 
the nonzero masses in equation~(\ref{m+}) receive one-loop corrections. 
However, $m_2 = \cdots = m_n$ is still valid because the unbroken symmetry 
group $G_{n-1}$ forbids such masses.

\paragraph{Case $\lambda' < 0$:}
For negative $\lambda'$, the condition 
\begin{equation}
|\lambda'| < \frac{n}{n-1} \lambda 
\end{equation}
is necessary for the scalar potential to be bounded from below. 
In this case the minimum is given by
\begin{equation}
v_1^2 = \cdots = v_n^2 = \frac{v^2}{n} \quad \Rightarrow \quad
v^{2}= \frac{-\mu^{2}}{\lambda + \frac{n-1}{n} \lambda'}.
\end{equation}
In principle, the VEVs $v_a$ could have different signs. However, since
arbitrary sign changes of the scalar fields are part of $G_n$, we can assume
$v_a > 0 \; \forall a$ without loss of generality.
Therefore, we have the symmetry breaking
\begin{equation}
G_n \xrightarrow{\text{SSB}} S_n,
\end{equation}
where the permutation group is given by its 
``natural permutation representation'' corresponding to $n \times n$ 
permutation matrices. 
This representation decays into the trivial 
one-dimensional and a $(n-1)$-dimensional irreducible representation.
Defining $n$ vectors $w_a$ ($a = 1,\ldots,n$) such that
\begin{equation}\label{w}
w_1 = \frac{1}{\sqrt{n}} \left(
\begin{array}{c} 1 \\ 1 \\ \vdots \\ 1 
\end{array} \right) 
\quad \mbox{and} \quad w_a \cdot w_b = \delta_{ab} \; \forall a,b,
\end{equation}
then $w_1$ is invariant under all permutation matrices and belongs, therefore,
to the trivial irreducible representation, while the vectors $w_2, \ldots w_n$ 
span the space pertaining to the $(n-1)$-dimensional one.
This is borne out by the tree-level masses. The scalars have the mass matrix
\begin{equation}
M^2_0 = A \mathbbm{1} + B w_1 w_1^T
\quad \mbox{with} \quad
A = -\frac{2 \lambda' v^2}{n}, \quad
B = 2 (\lambda + \lambda') v^2. 
\end{equation}
Hence, the diagonalization matrix is given by 
\begin{equation}\label{W0}
W_0 = \left( w_1, \ldots, w_n \right)
\end{equation}
and we find
\begin{equation}\label{M-}
M^2_{01} = A + B, \quad 
M^2_{02} = \cdots = M^2_{0n} = A.
\end{equation}
However, the fermion masses are all equal at tree level:
\begin{equation}\label{m-}
m_{01} = \cdots = m_{0n} =  \frac{yv}{\sqrt{n}}.
\end{equation}

At one-loop order, the scalar masses of equation~(\ref{M-}) will receive
radiative corrections, but---due to the unbroken symmetry group 
$S_n$---the relation $M^2_{2} = \cdots = M^2_{n}$ will still hold.

One might expect that the total degeneracy of the fermion masses, as expressed
in equation~(\ref{m-}), will be lifted because of radiative corrections such
that $m_1$ is different from the rest. However, as we will demonstrate now,
this is not the case. 

First we discuss the contribution from the finite one-loop 
VEVs shifts to the fermion masses. 
Since the fermion mass matrix is diagonal, we have
\begin{equation}\label{Yhat}
\hat Y_a = Y_b \left( W_0 \right)_{ba} = y \, \diag 
\left( \left( W_0 \right)_{1a}, \ldots, \left( W_0 \right)_{na} \right).
\end{equation}
In particular, 
\begin{equation}
\hat Y_1 = \frac{y}{\sqrt{n}}\,\mathbbm{1}
\quad \mbox{and} \quad
\mbox{Tr}\, \hat Y_a = 0 \quad \mbox{for}\; a = 2, \ldots, n
\end{equation}
due to $w_1$ of equation~(\ref{w}). Therefore, it follows from
equation~(\ref{Tchi}) that 
\begin{equation}
T^{(\chi)}_a = 0 \quad \mbox{for}\; a = 2, \ldots, n.
\end{equation}
Moreover, from equations~(\ref{vhat}) and~(\ref{W0}) we find
\begin{equation}
\hat v_1 = v, \quad \hat v_2 = \cdots = \hat v_n = 0.
\end{equation}
With this the tadpole expression $T^{(h)}_a$ of equation~(\ref{Th}) 
has the structure
\begin{equation}
T^{(h)}_a = \hat \lambda_{abrr} \hat v_b X_r = 
\hat \lambda_{a1rr} v X_r.
\end{equation}
According to equation~(\ref{eq:relation-lambda}), this expression can only be
nonzero for $a=1$. Therefore, 
\begin{equation}
T^{(h)}_a = 0 \quad \mbox{for}\; a = 2, \ldots, n
\end{equation}
as well and 
$\Delta \hat v_a \hat Y_a = \Delta \hat v_1 \hat Y_1 \propto \mathbbm{1}$.
This proves that the finite VEV shifts cannot remove the total fermion mass 
degeneracy.

Next we consider $\Sigma^{\textrm{1-loop}}$ of
equation~(\ref{sigma-1-loop}). We note that both $\hat D_a$ and $\hat E_a$ are
proportional to the unit matrix because of equation~(\ref{m-}). In addition,
because of equation~(\ref{M-}),
\begin{equation} 
\hat D_2 = \ldots = \hat D_n
\quad \mbox{and} \quad 
\hat E_2 = \ldots = \hat E_n.
\end{equation}
Thus we can write $\hat D_a = f_a \mathbbm{1}$ with
$f_2 = \cdots = f_n$,  but $f_1 \neq f_2$ in general. There are the analogous
relations for the $\hat E_a$. Considering now the $b$-th entry of the 
(diagonal) finite parts of $\Sigma^{\textrm{1-loop}}$ and taking into account
that the Yukawa coupling matrices are given by equation~(\ref{Yhat}), 
we have the generic sum 
\begin{equation}
\sum_{a=1}^n \left( W_0 \right)_{ba} f_a \left( W_0 \right)_{ba} = 
\left( W_0 \right)_{b1} \left( f_1 - f_2 \right) \left( W_0 \right)_{b1} + 
\sum_{a=1}^n \left( W_0 \right)_{ba} f_2 \left( W_0 \right)_{ba} = 
\frac{1}{n} \left( f_1 - f_2 \right) + f_2.
\end{equation}
(Note that there is no summation over the index $b$ in this equation.)
This result does not depend on $b$ and, therefore, $\Sigma^{\textrm{1-loop}}$ is 
proportional to the unit matrix. Consequently, the fermion mass degeneracy
cannot be lifted by one-loop contributions, as stated above.

\subsection{Soft symmetry breaking}
It is possible to lift any mass degeneracies by explicit breaking of $G_n$.
The model remains renormalizable, if we have soft breaking, for instance, 
by terms of dimension two. 
This is done by admitting in equation~(\ref{eq:relation-to-general}) a general
mass matrix $\mu^2_{ab}$, whereas the 
Yukawa and quartic couplings are 
still restricted by $G_n$.
This breaks the symmetry group $G_n$ down to
\begin{equation}
G \xrightarrow{\text{dim}\,2} (\mathbb{Z}_{4})_{\mathrm{diag}}
\end{equation}
with 
\begin{equation}
(\mathbb{Z}_{4})_{\mathrm{diag}}: \quad
\chi^\bb_{aL}\rightarrow i \chi^\bb_{aL}, \quad 
\varphi^\bb_{a} \rightarrow -\varphi^\bb_{a} \; \forall\,a,
\end{equation}
\textit{i.e.}\ this $\mathbb{Z}_{4}$ acts simultaneously on all fields and
agrees with equation~(\ref{S}). 
In this way, the scalar mass spectrum will be completely
non-degenerate already at tree level, but also the fermion mass spectrum
because a general matrix $\mu^2_{ab}$ will induce general VEVs $v_a$. 
It is easy to understand why this
modified model remains renormalizable; allowing for a general matrix 
$\mu^2_{ab}$, we also allow for a general counterterm matrix
$\delta\mu^2_{ab}$ and we can cancel the divergences related to the scalar
mass terms as handled by equation~(\ref{eq:delta-mu}).

It is natural that soft symmetry breaking is small.
We can easily incorporate this by taking one large mass parameter $\mu^2$ and
setting 
\begin{equation}
\mu^2_{ab} = \mu^2 \delta_{ab} + \sigma_{ab} 
\end{equation}
such that 
$\sum_{a=1}^n \sigma_{aa} = 0$ and 
$|\sigma_{ab}| \ll \mu^{2}$ $\forall\, a,b$.
In this case the previously degenerate masses will now become slightly
different and we can produce quasi-degenerate mass spectra.

\section{Dirac fermions}
\label{dirac}
So far, we have put the focus on Majorana fermions. We have done so because 
in the long run we are interested in studying radiative corrections 
in neutrino mass models which typically feature the seesaw mechanism and,
therefore, neutrinos of Majorana nature.
However, it is straightforward to switch from Majorana to 
Dirac fermions. How this is done will be explained in this section---see 
also~\cite{Grimus:2014zwa,Denner:1992me}.

\paragraph{Lagrangian, diagonalization of Dirac mass matrices, and 
renormalization:}
In the Dirac setup, we can in general have $n_{\chi_L}$ chiral 
fields $\chi^\bb_{iL}$ and $n_{\chi_R}$ independent chiral fields $\chi^\bb_{iR}$, 
while the scalar sector remains the same as in the Majorana case. 
Then, the bare Lagrangian is given by
\begin{subequations}\label{eq:L-Dirac}
\begin{eqnarray}
\mathcal{L}_{B} &=& 
i \bar \chi^\bb_{iL} \gamma^\mu \partial_\mu \chi^\bb_{iL} +
i \bar \chi^\bb_{iR} \gamma^\mu \partial_\mu \chi^\bb_{iR} +
\frac{1}{2} \left( \partial_\mu \varphi^\bb_a \right) 
\left( \partial^\mu \varphi^\bb_a \right)
\\ &&
- \left(\left(Y^\bb_a\right)_{ij}\, 
{\bar{\chi}^\bb_{iR}} \chi^\bb_{jL} \varphi^\bb_a + \mbox{H.c.} \right) 
\label{YD} \\ &&
- \frac{1}{2} ( \mu_{\scriptscriptstyle B}^2 )_{ab} \varphi^\bb_a \varphi^\bb_b -
\frac{1}{4} \lambda^\bb_{abcd}\, 
\varphi^\bb_a \varphi^\bb_b \varphi^\bb_c \varphi^\bb_d,
\end{eqnarray}
\end{subequations}
where the $Y^\bb_a$ now are $n_\varphi$ general complex 
$n_{\chi_R} \times n_{\chi_L}$ matrices.
In principle, $n_{\chi_L}$ could be different from $n_{\chi_R}$, in which case one
has $|n_{\chi_L} - n_{\chi_R}|$ massless Weyl fermions. However, for simplicity
we assume $n_{\chi_L} = n_{\chi_R} \equiv n_\chi$ in the following. 
A possible modification of the transformation of the fermions in 
equation~(\ref{S}) is the $\mathbbm{Z}_2$ symmetry
\begin{equation}
\mathcal{S}': \quad \chi^\bb_L \rightarrow -\chi^\bb_L, \quad
\chi^\bb_R \rightarrow \chi^\bb_R, \quad \varphi^\bb \to -\varphi^\bb,
\end{equation}
in order to forbid fermion tree-level mass terms and linear and 
trilinear terms in the scalar potential.

The renormalization of the fermionic fields now becomes
\begin{equation}
 \chi^\bb_{L} = Z_{\chi_L}^{(1/2)}\, \chi_{L}, \quad
 \chi^\bb_{R} = Z_{\chi_R}^{(1/2)}\, \chi_{R},
\end{equation}
involving two independent general complex matrices 
$Z_{\chi_L}^{(1/2)}$ and $Z_{\chi_R}^{(1/2)}$.
Inserting this into equation~(\ref{eq:L-Dirac}) yields a renormalized 
Lagrangian with Yukawa coupling matrices $Y_a$ 
and counterterms similar to the Majorana case.
The main changes lie in the definition of the Yukawa counterterm
\begin{equation}
 \mc^{\varepsilon/2} \delta Y_a =
\left( Z_{\chi_R}^{(1/2)} \right)^\dagger Y^\bb_b\, Z_{\chi_L}^{(1/2)} 
\left( Z^{(1/2)}_\varphi \right)_{ba} - \mc^{\varepsilon/2} Y_a,
\end{equation}
and the need for the definition of two independent hermitian matrices
\begin{equation}
\delta^{(\chi_L)} = 
\left( Z_{\chi_L}^{(1/2)} \right)^\dagger Z_{\chi_L}^{(1/2)} - \bone, \quad
\delta^{(\chi_R)} = 
\left( Z_{\chi_R}^{(1/2)} \right)^\dagger Z_{\chi_R}^{(1/2)} - \bone.
\end{equation}

Via SSB we obtain the tree-level Dirac mass matrix 
\begin{equation}
m_0 = \sum_{a=1}^{n_\varphi} v_a Y_a.
\end{equation}
This mass matrix is bi-diagonalized with two unitary matrices 
$U_{L0}$ and $U_{R0}$:
\begin{equation}
U_{R0}^\dagger m_0 U_{L0} = \hat{m}_0 \equiv 
\diag \left( m_{01}, \ldots, m_{0 n_{\chi}} \right).
\end{equation}
Due to the left and right diagonalization matrices, there are now left 
and right chiral mass eigenfields 
\begin{equation}
\hat{\chi}_{L} = U_{L0}^\dagger \chi_{L}, \quad 
\hat{\chi}_{R} = U_{R0}^\dagger \chi_{R}.
\end{equation}
Moreover, equations~(\ref{eq:hat-delta-chi}) and~(\ref{eq:hat-Y}) 
are modified to
\begin{subequations}
\begin{alignat}{2}
\delta^{(\chi_L)} &\rightarrow \hat{\delta}^{(\chi_L)} = 
    U_{L0}^\dagger \delta^{(\chi_L)} U_{L0}, &
    \delta^{(\chi_R)} &\rightarrow \hat{\delta}^{(\chi_R)} = 
    U_{R0}^\dagger \delta^{(\chi_R)} U_{R0}, \\
Y_a &\rightarrow \hat{Y}_a = \left(U_{R0}^\dagger  Y_{b} U_{L0} \right) (W_0)_{ba},
\end{alignat}
\end{subequations}
respectively.

In analogy to equation~(\ref{mef-m}), we define Dirac mass eigenfields 
\begin{equation}
\hat\chi_i = \hat\chi_{iL} + \hat\chi_{iR}
\end{equation}
and the corresponding vector of eigenfields $\hat\chi$.
In terms of mass eigenfields, the Yukawa interaction reads
\begin{equation}\label{Y-D} 
\mathcal{L}_Y = - \bar{\hat\chi} 
\left( \hat Y_a \gamma_L + {\hat Y}^\dagger_a  \gamma_R \right) \hat\chi 
\left( \mc^{\varepsilon/2} \hat h_a + \hat{\bar v}_a \right).
\end{equation}
Formally, Dirac and Majorana Yukawa terms look the same~\cite{Denner:1992me}.
Note that the only difference of this $\mathcal{L}_Y$ to that of 
equation~(\ref{Y}) is the factor $1/2$ which we do not introduce 
in the Dirac case. It will become clear in 
the last paragraph of this section why we prefer this definition.

\paragraph{Fermion selfenergy:}
With the above definitions, the renormalization programme of 
section~\ref{renormalization} goes through with only minor modifications.
The renormalized fermion selfenergy for Dirac fermions is given by
\begin{eqnarray}
\label{Sigma-D}
\Sigma(p) &=&
\Sigma^{\textrm{1-loop}}(p) 
- \slashed{p} \left[ \hat{\delta}^{(\chi_L)} \gamma_L + 
\hat{\delta}^{(\chi_R)} \gamma_R \right]
\nonumber \\ && 
+ \hat{v}_a \left[ \delta {\hat{Y}}_a \gamma_L + 
\left(\delta {\hat{Y}}_a\right)^\dagger \gamma_R \right] + 
\Delta \hat{v}_a \left[ {\hat{Y}}_a \gamma_L + 
\hat{Y}_a^\dagger \gamma_R \right].
\end{eqnarray}
Eventually, the one-loop Dirac masses read
\begin{eqnarray}
m_i &=& m_{0i} + \frac{1}{2} m_{0i} \left[ 
\left( \Sigma^{(A)}_L \right)_{ii} (m_{0i}^2) + 
\left( \Sigma^{(A)}_R \right)_{ii} (m_{0i}^2) \right] +
\mbox{Re} \left( \Sigma^{(B)}_L \right)_{ii}(m^2_{0i}).
\end{eqnarray}
Note that 
$ \mbox{Re} \left( \Sigma^{(B)}_L \right)_{ii} = 
\mbox{Re}  \left( \Sigma^{(B)}_R \right)_{ii}$ 
because of the symmetry relation~(\ref{sigma-h}).

\paragraph{Computation of amplitudes:} 
There are two changes when we switch from 
Majorana to Dirac fermions~\cite{Grimus:2014zwa}:
\begin{enumerate}
\renewcommand{\labelenumi}{\roman{enumi}.}
\item
$\hat{Y}_a^* \to \hat{Y}_a^\dagger$ and 
\item
a factor of two for every closed Dirac fermion loop compared to the  
corresponding Majorana fermion loop.
\end{enumerate}
As discussed above, the first change simply comes from the fact that for 
Dirac neutrinos the Yukawa coupling matrices are not symmetric. 
The reason for the factor of two is the following. 
In the Majorana case we have defined the Yukawa Lagrangian with a 
factor $1/2$---see equation~(\ref{Y}).
If a Majorana fermion line in a Feynman diagram is not closed, then all 
factors of $1/2$ are cancelled because, whenever a fermion line 
is connected to a vertex, there are two possible Wick contractions;
however, in a closed loop one factor $1/2$ is left over because,
when closing the loop, there is only one contraction.
In the Dirac case, we have omitted the factor $1/2$ in the 
Yukawa Lagrangian~(\ref{Y-D}) because, when we connect a Dirac fermion 
line to a vertex, there is exactly one Wick contraction. 
Therefore, when a closed fermion loop occurs, there is a factor 
of two for Dirac fermions relative to Majorana fermions.
Finally, whenever we have made a simplification in a trace by exploiting 
$Y_a^T = Y_a$ in the Majorana case, as done in equations~(\ref{Pi-a}) 
and~(\ref{dlch}), we have to revoke it in the Dirac case.

Consequently, in the Dirac case, $\Pi^{(a)}(p^2)$ is given by 
\begin{eqnarray}
\Pi^{(a)}_{ab}(p^2) &=&
\frac{2}{16 \pi^2} \left\{ c_\infty \,\mbox{Tr} \left[
{\hat{Y}}_a {\hat{m}_0} {\hat{Y}}_b {\hat{m}_0} + 
{\hat{Y}}_a^\dagger \hat{m}_0 {\hat{Y}}_b^\dagger \hat{m}_0 + 
{\hat{Y}}_a {\hat{Y}}_b^\dagger {\hat{m}_0}^2  + 
{\hat{Y}}_a^\dagger {\hat{Y}}_b {\hat{m}_0}^2 \right. \right.
\nonumber \\ && 
\left. + 
{\hat{Y}}_a {\hat{m}_0}^2 {\hat{Y}}_b^\dagger  + 
{\hat{Y}}_a^\dagger {\hat{m}_0}^2 {\hat{Y}}_b 
\right] 
-\frac{1}{2} c_\infty \, \mbox{Tr} \left[ {\hat{Y}}_a {\hat{Y}}_b^\dagger + 
{\hat{Y}}_a^\dagger {\hat{Y}}_b \right] p^2
\nonumber \\ && \left. +\frac{1}{2} \,\mbox{Tr} \left[
\left( \hat{Y}_a \hat{Y}_b^\dagger + \hat{Y}_b^\dagger \hat{Y}_a +  
\hat{Y}_a^\dagger \hat{Y}_b + \hat{Y}_b \hat{Y}_a^\dagger \right)
\left( \hat{m}_0^2 - \frac{1}{6}\, p^2 \right) \right]
- \cdots \right\}.
\label{Pi-a-d}
\end{eqnarray}
The dots refer to the integral in equation~(\ref{Pi-a}) 
where merely $Y_a^*$ has to be substituted 
by $Y_a^\dagger$. From equation~(\ref{Pi-a-d}),
$\big( \Pi^{(a)}_\infty \big)_{ab}$ can be read off.
Equation~(\ref{dlch}) is modified to 
\begin{equation}
\delta{\hat{\lambda}}^{(\chi)}_{abcd} = -\frac{1}{3} \times
\frac{1}{16 \pi^2} c_\infty 
\mathrm{Tr} \left[ 
{\hat{Y}}_a {\hat{Y}}_b^\dagger {\hat{Y}}_c {\hat{Y}}_d^\dagger + 
\cdots + 
{\hat{Y}}_a^\dagger {\hat{Y}}_b {\hat{Y}}_c^\dagger {\hat{Y}}_d + 
\cdots \right],
\label{dlch-d}
\end{equation}
where the dots indicate the five non-trivial permutations of the indices
$b,c,d$.
No complications arise in 
equations~(\ref{Tchi}), (\ref{c-delta-h}), (\ref{c1}) and~(\ref{c2});
for Dirac fermions 
one simply has to multiply the right-hand side by a factor of two and
replace complex conjugation by hermitian conjugation.

\section{Conclusions}
\label{conclusions}
In this paper we have presented a versatile and simple renormalization
procedure which is adapted to models which have SSB and a multitude of
scalars. 
This renormalization programme takes seriously 
the nature of masses as functions of the parameters of 
the underlying model; therefore, physical masses have an expansion 
in perturbation theory just like any other observable. 
We have exemplified our renormalization procedure by discussing  
a general Yukawa model with an arbitrary number of fermion fields 
of Majorana or Dirac nature and an arbitrary number of real scalar
fields; moreover, this toy model 
has the feature that tree-level fermion masses
are generated by SSB of a cyclic group. 
In particular, we have explicitly computed the fermionic and scalar 
selfenergies and studied radiative corrections at the one-loop level to
tree-level masses.

The main idea discussed in this paper is to split renormalization into a step
in which UV divergent parts are cancelled by $\overline{\mbox{MS}}$
renormalization of the parameters of the unbroken theory
and a subsequent step in which finite corrections are performed to make the
scalar one-point functions vanish and to obtain one-loop pole masses.
We have presented the details of the cancellation of UV divergences and 
elucidated the role of tadpole diagrams in our renormalization procedure 
and their contributions to the masses.
We have also applied our findings to a showcase model furnished with a 
non-Abelian flavour symmetry group.

A typical example where the renormalization procedure put forward in this
paper can be applied is the lepton 
sector of the multi-Higgs-doublet Standard Model with 
an arbitrary number of right-handed neutrino singlets 
and flavour symmetries; this comprises the seesaw mechanism as well as 
light sterile neutrinos. A derivation of 
general formulae which permit to compute radiative corrections to tree-level 
predictions of masses and mixing angles in this rather general class of 
flavour models is in preparation.

\section*{Acknowledgments}
M.L. is supported by the Austrian Science Fund (FWF), Project
No.\ P28085-N27 and in part by the FWF Doctoral Program No.\ W1252-N27 
Particles and Interactions. 
The authors thank H. Eberl, G. Ecker, M. M\"uhlleitner and H.~Neufeld 
for stimulating discussions. M.L.\ also thanks D. Lechner and C. Lepenik 
for further helpful discussions.

\newpage

\appendix

\section{Selfenergies and on-shell renormalization}
\setcounter{equation}{0}
\renewcommand{\theequation}{A\arabic{equation}}
\label{selfenergies}
Since for fermions the general relations
\begin{equation}\label{sigma-h}
\left( \Sigma^{(A)}_L \right)^\dagger = \Sigma^{(A)}_L, \quad
\left( \Sigma^{(A)}_R \right)^\dagger = \Sigma^{(A)}_R, \quad
\left( \Sigma^{(B)}_L \right)^\dagger = \Sigma^{(B)}_R 
\end{equation}
are valid,\footnote{Strictly speaking these relations hold only for the
  dispersive part of the selfenergy.}
we see that for the finiteness of $\Sigma^{(A)}_L$ and $\Sigma^{(A)}_R$
the counterterm with the hermitian $\delta^{(\chi)}$ suffices.
In addition, we remark that in the case of Majorana fermions the further
conditions~\cite{Denner:1992me,lavoura}
\begin{equation}\label{sigma-t}
\left( \Sigma^{(A)}_L \right)^T = \Sigma^{(A)}_R, \quad
\left( \Sigma^{(B)}_L \right)^T = \Sigma^{(B)}_L, \quad
\left( \Sigma^{(B)}_R \right)^T = \Sigma^{(B)}_R
\end{equation}
hold. This is a general condition, but can also be seen explicitly in 
our one-loop result.

In order to switch from the renormalized Majorana selfenergy 
$\Sigma(p)$ and the bosonic selfenergy $\Pi(p^2)$ 
to the on-shell selfenergies $\widetilde\Sigma(p)$ and $\widetilde\Pi(p^2)$, 
respectively, we must allow for finite field strength 
renormalization matrices. Denoting these by 
\begin{equation}
\overset{\circ}{Z}^{\raisebox{-6pt}{$\scriptstyle (1/2)$}}_\chi = \bone + 
\frac{1}{2}\overset{\circ}{z}_\chi 
\quad \mbox{and } \quad
\overset{\circ}{Z}^{\raisebox{-6pt}{$\scriptstyle (1/2)$}}_h = \bone + 
\frac{1}{2}\overset{\circ}{z}_h,
\end{equation}
we have at one-loop order
\begin{eqnarray}
\widetilde\Sigma(p) &=& 
\Sigma(p) - \frac{1}{2}\, \slashed{p} \left[
\left( \Big( \overset{\circ}{z}_\chi \Big)^\dagger + 
\overset{\circ}{z}_\chi \right) \gamma_L + 
\left( \Big( \overset{\circ}{z}_\chi \Big)^\dagger + 
\overset{\circ}{z}_\chi \right)^* \gamma_R \right]
\nonumber \\ && +
\frac{1}{2} \left[ \left( \Big (\overset{\circ}{z}_\chi \Big)^T \hat{m}_0 + 
\hat{m}_0\, \overset{\circ}{z}_\chi \right) \gamma_L + 
\left( \Big(\overset{\circ}{z}_\chi \Big)^T \hat{m}_0 + 
\hat{m}_0\, \overset{\circ}{z}_\chi \right)^* \gamma_R \right],
\\
\widetilde\Pi_{ab}(p^2) &=& \Pi_{ab}(p^2) - 
\frac{1}{2} \left[ \left( \overset{\circ}{z}_h  \right)^T + 
\overset{\circ}{z}_h  \right]_{ab} p^2
+ \frac{1}{2} \left[ \left( \overset{\circ}{z}_h  \right)^T {\hat M}^2_0
+ {\hat M}^2_0\, \overset{\circ}{z}_h  \right]_{ab}.
\end{eqnarray}
It is important to note that we have no freedom for mass renormalization 
because in our scheme the masses are computed in terms of the renormalized 
parameters of the model. 
Due to the Majorana nature of the fermions under consideration,
the relation
\begin{equation}
\overset{\circ}{z}_\chi \equiv 
\left( \overset{\circ}{z}_L \right)_{ij} = 
\left( \overset{\circ}{z}_R \right)_{ij}^*
\end{equation}
holds for left and right-chiral fields. 
In $\widetilde\Sigma(p)$ this fact has been taken into account.
Using the second relation in equation~(\ref{sigma-h}) and 
the first relation in equation~(\ref{sigma-t}), the on-shell conditions
lead for $i \neq j$ to~\cite{Denner:1990,kiyoura,Grimus:2016hmw,grimus}
\begin{eqnarray}\label{zL}
\lefteqn{\frac{1}{2}(\overset{\circ}{z}_\chi)_{ij} =} \\
&& -\frac{1}{m_{0i}^2-m_{0j}^2} 
\left[ m_{0j}^2 \left( \Sigma^{(A)}_L \right)_{ij} + 
m_{0i} m_{0j} \left( \Sigma^{(A)}_L \right)_{ji} + 
m_{0j} \left( \Sigma^{(B)}_L \right)_{ji}^* +
m_{0i} \left( \Sigma^{(B)}_L \right)_{ij} \right]_{p^2 = m_{0j}^2}.
\nonumber
\end{eqnarray}
Furthermore, for $i = j$ we obtain
\begin{eqnarray}
\lefteqn{\mathrm{Re}(\overset{\circ}{z}_\chi)_{ii} =} 
\\ && \nonumber
(\Sigma^{(A)}_L)_{ii}(m_{0i}^2) + 2m_{0i}^2
\left. \frac{\dd}{\dd p^2}
\left( \Sigma^{(A)}_L)_{ii}(p^2) \right)\right|_{p^2=m_{0i}^2} + 
2m_{0i} \left. \frac{\dd}{\dd p^2}\;\mbox{Re} 
\left( \Sigma^{(B)}_L)_{ii}(p^2) \right)\right|_{p^2=m_{0i}^2}
\end{eqnarray}
and
\begin{equation}\label{Imzii}
m_{0i}\, \mathrm{Im}\, (\overset{\circ}{z}_\chi)_{ii} = 
-\mathrm{Im}\,(\Sigma^{(B)}_L)_{ii}(m_{0i}^2).
\end{equation}
It is characteristic of Majorana fermions that there is no phase freedom
in the determination of the field strength renormalization matrix,
\textit{i.e.}\ not only the real part but also the imaginary part of 
$(z_\chi)_{ii}$ is fixed.

Finally, in the scalar scalar case we are lead to 
\begin{equation}
a \neq b\!: \; \frac{1}{2} \left( \overset{\circ}{z}_h  \right)_{ab} = 
-\frac{\Pi_{ab}(M_{0b}^2)}{M_{0a}^2 - M_{0b}^2}, \quad
a = b\!: \; \left( \overset{\circ}{z}_h  \right)_{aa} = 
\left. \frac{\dd \Pi_{aa}(p^2)}{\dd p^2} \right|_{p^2 =  M_{0a}^2}
\end{equation}
for on-shell renormalization.

\section{Finite tadpole contributions}
\setcounter{equation}{0}
\renewcommand{\theequation}{B\arabic{equation}}
\label{finite tadpole}
Throughout this appendix the discussion refers to the one-loop order.
In the fermionic as well as the scalar selfenergy, 
tadpole diagrams contribute indirectly via the 
finite shift~(\ref{eq:vev-shift}), even though in both cases 
the condition $t_a = 0$ of equation~(\ref{eq:tadpole-condition}) 
and the requirement that the scalar one-point function is zero---see 
equation~(\ref{eq:tadpole-condition-graph})---procure
the vanishing of the sum of tadpole diagrams and the term
\begin{equation}\label{tct}
\Delta \hat{t}_a + \delta\hat{\mu}^2_{ab} \hat{v}_b + 
\delta\hat{\lambda}_{abcd}\, \hat{v}_b \hat{v}_c \hat{v}_d. 
\end{equation}
Diagrammatically, this can be written as 
\begin{equation}
  \begin{fmffile}{fermion-two-point-tadpoles}
    \begin{gathered}
      \begin{fmfgraph*}(50,40)
	\fmftop{i1}
	\fmfbottom{b1,b2}
	\fmf{plain}{b1,v2,b2}
	\fmffreeze
	\fmf{phantom}{i1,v1,v2}
	\fmf{dashes}{v2,v1}
	\fmf{plain,left}{v1,i1,v1}
      \end{fmfgraph*}
    \end{gathered}
    + 
    \begin{gathered}
      \begin{fmfgraph*}(50,40)
	\fmftop{i1}
	\fmfbottom{b1,b2}
	\fmf{plain}{b1,v2,b2}
	\fmffreeze
	\fmf{phantom}{i1,v1,v2}
	\fmf{dashes}{v2,v1}
	\fmf{dashes,left}{v1,i1,v1}
      \end{fmfgraph*}
    \end{gathered}
    + \begin{gathered}
      \begin{fmfgraph*}(50,40)
	\fmftop{i1}
	\fmfbottom{b1,b2}
	\fmf{plain}{b1,v2,b2}
	\fmffreeze
	\fmf{phantom}{i1,v1,v2}
	\fmf{dashes}{v2,v1}
	\fmf{phantom,left}{v1,i1,v1}
	\fmfv{decor.shape=cross}{v1}
      \end{fmfgraph*}
    \end{gathered}
    = 0
  \end{fmffile} 
\end{equation}
and
\begin{equation}
  \begin{fmffile}{scalar-two-point-tadpoles}
    \begin{gathered}
      \begin{fmfgraph*}(50,40)
	\fmftop{i1}
	\fmfbottom{b1,b2}
	\fmf{dashes}{b1,v2,b2}
	\fmffreeze
	\fmf{phantom}{i1,v1,v2}
	\fmf{dashes}{v2,v1}
	\fmf{plain,left}{v1,i1,v1}
      \end{fmfgraph*}
    \end{gathered}
    + 
    \begin{gathered}
      \begin{fmfgraph*}(50,40)
	\fmftop{i1}
	\fmfbottom{b1,b2}
	\fmf{dashes}{b1,v2,b2}
	\fmffreeze
	\fmf{phantom}{i1,v1,v2}
	\fmf{dashes}{v2,v1}
	\fmf{dashes,left}{v1,i1,v1}
      \end{fmfgraph*}
    \end{gathered}
    + \begin{gathered}
      \begin{fmfgraph*}(50,40)
	\fmftop{i1}
	\fmfbottom{b1,b2}
	\fmf{dashes}{b1,v2,b2}
	\fmffreeze
	\fmf{phantom}{i1,v1,v2}
	\fmf{dashes}{v2,v1}
	\fmf{phantom,left}{v1,i1,v1}
	\fmfv{decor.shape=cross}{v1}
      \end{fmfgraph*}
    \end{gathered}
    = 0,
  \end{fmffile} 
\end{equation}
where the cross symbolizes the contribution of equation~(\ref{tct}).
Still, the finite parts of the tadpole diagrams generate, via the finite 
VEV shifts $\Delta v_a$,  
the mass shifts
\begin{equation}
 \Delta \hat{m}_0 = \hat{Y}_a \Delta \hat{v}_a 
\end{equation}
for the fermions---see equation~(\ref{Sigma})---and
\begin{equation}
(\Delta \hat{M}_0^2)_{ab} = 6 \hat{\lambda}_{abcd} \hat v_c \Delta \hat{v}_d
\end{equation}
for the real scalars---see equation~(\ref{Pi}). 
These add to the counterterms of the 
fermionic and scalar two-point functions.
In terms of diagrams, this can be symbolized as 
\begin{equation}
  \begin{fmffile}{counterterm-fermion}
    \begin{gathered}
      \begin{fmfgraph*}(50,40)
	\fmfleft{i}
	\fmfright{o}
	\fmf{plain}{i,v,o}
	\fmfv{decor.shape=cross}{v}
      \end{fmfgraph*}
    \end{gathered}
    =  -i\left( \delta \hat{Y_a} \hat{v}_a + \Delta \hat{m}_0 \right)
  \end{fmffile} 
\end{equation}
for the fermions and 
\begin{equation}
  \begin{fmffile}{counterterm-scalar}
    \begin{gathered}
      \begin{fmfgraph*}(50,40)
	\fmfleft{i}
	\fmfright{o}
	\fmf{dashes}{i,v,o}
	\fmfv{decor.shape=cross}{v}
      \end{fmfgraph*}
    \end{gathered}
    =  -i \left( \delta \hat{\mu}^2_{ab} + 
      3 \delta \hat{\lambda}_{abcd}\hat{v}_c \hat{v}_d
      + (\Delta \hat{M}^2_0)_{ab} \right) 
  \end{fmffile} 
\end{equation}
for the scalars.

\section{Integrals}
\setcounter{equation}{0}
\renewcommand{\theequation}{C\arabic{equation}}
\label{integrals}
\begin{eqnarray}
\mc^\varepsilon \int \frac{\mathrm{d}^d k}{(2\pi)^d} \,
\frac{1}{ k^2 - \Delta + i\epsilon } 
&=&
\frac{i}{16\pi^2} \, \Delta \left( c_\infty + 1 - 
\ln \frac{\Delta}{\mc^2} \right),  
\\
\mc^\varepsilon \int \frac{\mathrm{d}^d k}{(2\pi)^d} \,
\frac{1}{ ( k^2 - \Delta + i\epsilon )^2} 
&=&
\frac{i}{16\pi^2} \left( c_\infty - \ln \frac{\Delta}{\mc^2} \right),
\\
\mc^\varepsilon \int \frac{\mathrm{d}^d k}{(2\pi)^d} \,
\frac{k^2}{ ( k^2 - \Delta + i\epsilon )^2} 
&=&
\frac{i}{16\pi^2} \, \Delta 
\left( 2c_\infty + 1 - 2\ln \frac{\Delta}{\mc^2} \right).
\end{eqnarray}

\end{document}